\documentclass[12pt,preprint]{aastex}
\usepackage{eqnarray}
\usepackage{natbib}
\usepackage{threeparttable}
\usepackage{rotating}
\usepackage{lscape}
\shortauthors{Kryukova et al.}

\bibpunct{(}{)}{,}{a}{}{;}
\bibpunct{(}{)}{,}{a}{}{;}

\begin{document}

\pdfoutput=1

\title{Luminosity Functions of Spitzer Identified Protostars in Nine Nearby Molecular Clouds}

\author{E. Kryukova\altaffilmark{1}; S.~T. Megeath\altaffilmark{1}, R.~A. Gutermuth\altaffilmark{2}, J. Pipher\altaffilmark{3}, T.~S. Allen\altaffilmark{1}, L.~E. Allen\altaffilmark{4}, P.~C. Myers\altaffilmark{5}, J. Muzerolle\altaffilmark{6}}

\altaffiltext{1}{Department of Physics and Astronomy, University of Toledo, Toledo, OH}
\altaffiltext{2}{Department of Astronomy, University of Massachusetts, Amherst, MA}
\altaffiltext{3}{Department of Physics and Astronomy, University of Rochester, Rochester, NY}
\altaffiltext{4}{National Optical Astronomy Observatories, Tucson, AZ}
\altaffiltext{5}{Harvard-Smithsonian Center for Astrophysics, Cambridge, MA}
\altaffiltext{6}{Space Telescope Science Institute, Baltimore, MD}

\begin{abstract}
We identify protostars in {\it Spitzer} surveys of nine star-forming molecular clouds within 1 kpc: Serpens, Perseus, Ophiuchus, Chamaeleon, Lupus, Taurus, Orion, Cep OB3, and Mon R2, which combined host over 700 protostar candidates.  These clouds encompass a variety of star forming environments, including both low mass and high mass star forming regions, as well as dense clusters and regions of sparsely distributed star formation.  Our diverse cloud sample allows us to compare protostar luminosity functions in these varied environments.  We combine near- and mid-infrared photometry from 2MASS J, H, and $K_{s}$ bands and {\it Spitzer} 3.6, 4.5, 5.8, 8.0, and 24~$\mu$m bands to create 1 - 24~$\mu$m spectral energy distributions (SEDs).  Using protostars from the {\it c2d} survey with well-determined bolometric luminosities, we derive a relationship between bolometric luminosity, mid-IR luminosity (integrated from 1 - 24~$\mu$m), and SED slope.  Estimations of the bolometric luminosities for protostar candidates are combined to create luminosity functions for each cloud.  Contamination due to edge-on disks, reddened Class II sources, and galaxies is estimated and removed from the luminosity functions.  We find that luminosity functions for high mass star forming clouds (Orion, Mon R2, and Cep OB3) peak near 1~L$_{\odot}$ and show a tail extending toward luminosities above 100~L$_{\odot}$.  The luminosity functions of the low mass star forming clouds (Serpens, Perseus, Ophiuchus, Taurus, Lupus, and Chamaeleon) do not exhibit a common peak, however the combined luminosity function of these regions peaks below 1~L$_{\odot}$.  Finally, we examine the luminosity functions as a function of the local surface density of YSOs.  In the Orion molecular cloud, we find a significant difference between the luminosity functions of protostars in regions of high and low stellar density, the former of which is biased toward more luminous sources.  This may be the result of primordial mass segregation, although this interpretation is not unique.  We compare our luminosity functions to those predicted by models and find that our observed luminosity functions are best matched by models which invoke competitive accretion, although we do not find strong agreement of the high mass star forming clouds with any of the models.

\end{abstract}


\section{Introduction} 

Stars form in a diverse range of environments within molecular clouds.   These cloud environments include crowded, massive clusters heated by O stars (e.g. in Orion), smaller clusters without high mass stars (e.g. in Ophiuchus), or isolated, cold dark clouds containing low mass stars (e.g. in Taurus).  The temperatures, densities, and turbulent linewidths of the natal molecular gas can vary systematically between these regions \citep{1999ApJS..125..161J,1999ApJ...525..343W}, and the surface density of young stellar objects (YSOs) can vary by over two orders of magnitude \citep{2011ApJ...739...84G}. This motivates us to study how these different environments affect the outcomes of the star formation process.

Despite significant differences in gas temperature, column density, and the turbulent velocities, the star forming regions and clusters in our galaxy exhibit remarkably similar initial mass functions \citep[hereafter: IMFs,][]{2010ARA&A..48..339B}.  There may be a few exceptions: \cite{2009ApJ...703..399L} have compared the IMFs from Taurus with Chamaeleon~I and IC-348 and find that Taurus has a significant excess of stars between 0.6 and 0.8 M$_{\odot}$.  Nevertheless, the similarity between the IMFs of dark clouds like Chamaeleon~I, clusters with B-stars such as IC~348, and massive clusters with O and B stars such as the Orion Nebula Cluster suggests that the IMF is remarkably invariant.  Although the IMF averaged over a star forming region or cluster may be universal, there is some evidence that masses are initially segregated within clouds or within clusters.  Massive stars are found primarily in the center of clusters, although it not clear whether this is the result of primordial mass segregation or dynamical evolution \citep{1998MNRAS.295..691B,2009MNRAS.400..657M}.  The detection of compact groups of massive protostars in the centers of clusters is evidence for primordial mass segregation since such objects would have little time for dynamical evolution \citep{2005ApJ...622L.141M,2006ApJ...649..888H}.   Mass segregation is also observed in small groups of young stars: \cite{2011ApJ...727...64K} studied groups of 20 and 40 stars in the Taurus and Perseus molecular clouds and found that the most massive members (in some cases with a mass of only 1 M$_{\odot}$) are found preferentially near the centers of the groups.  In such small groups, it is unlikely that dynamical evolution has occurred since the relaxation time of the group exceeds 5 Myr, longer than the estimated ages of the groups \citep{2011ApJ...727...64K}.  These observations are evidence for primordial mass segregation, with the IMF in the centers of clusters biased towards higher mass stars.

An understanding of the origin of the IMF, and the potential primordial mass segregation within star forming regions, requires a better knowledge of both the initial conditions leading to the formation of stars as well as the process by which the gas is subsequently accreted onto the star.  Work on the initial conditions has focused on the determining the mass function of dense molecular cores that collapse into stars (hereafter: core mass function or CMF).  The similarity in the shapes of the IMF and CMF has led to the suggestion that the IMF reflects the CMF at later stages if a fixed percentage of the core mass is accreted onto the central protostar \citep[e.g.][]{2007A&A...462L..17A,2010A&A...518L.102A}.  Although this provides an attractive model for a universal IMF, there are several questions regarding this model.  First, completeness and signal-to-noise issues may contribute to the observed shape of the CMF \citep{2010ApJ...719..561R}.  Second, it is unclear whether many of the observed star-less cores are gravitationally bound and capable of forming stars \citep{2008ApJ...672..410L}.  Third, the most massive cores are observed to form groups of stars, not single stars, and consequently, the high mass end of the CMF may describe a mass function for small stellar groups and not individual stars \citep{2007ApJ...655..364B,2009ApJ...704..183R,2008ApJ...679..552S}.  Fourth, the collapse of the core may be followed by the infall of gas from the surrounding molecular cloud resulting in a final stellar mass greater than that of the initial core \citep{2009ApJ...706.1341M}.  Finally, it is known that young stars have stellar outflows and winds that mediate the infall of gas onto the central system and the accretion of gas onto the central protostar, thus the final mass must be in part determined by protostellar evolution \citep{1996ApJ...464..256A}.  Like the IMF, a number of investigations have found that the CMF is also remarkably invariant in star forming clouds \citep[e.g.][]{2010ApJ...710.1247S}.   However, these studies may not be able to detect variations in the CMF due to the limited angular resolution of the observations and the relatively small number of detected cores.  Furthermore, it remains to be seen if the core masses are segregated within molecular clouds.

The study presented in this paper is part of a larger effort to build a bridge between the initial conditions of star formation (the CMF) and the resulting ensemble properties of the nascent stars (the IMF).   It is in the protostellar phase that the mass from the collapsing core is accreted onto a star; thus, protostellar evolution must play a key role in determining the  properties of a nascent star.  Our understanding of protostars is limited by the difficulty in measuring their basic properties.  The radiation from the accreting protostar is reprocessed or scattered by a non-spherical infalling envelope and disk.  Consequently, observations spanning near- to far-IR wavelength coupled with radiative transfer codes are needed to constrain fundamental attributes  such as the emitted protostar luminosity, the  inner envelope density, and the geometry of the envelope and outflow cavities. For this reason, current work has focused on the most readily measured property of a protostar: its observed luminosity.  The determination of the protostellar luminosity function in the nearest clouds by the {\it Spitzer} c2d legacy program has motivated several theoretical studies showing that the luminosity functions provide an important constraint on protostellar evolution \citep{2010ApJ...710..470D,2011ApJ...736...53O,2011ApJ...743...98M}.   

We present here an observational study of protostellar luminosity functions in nine nearby ($<$ 1 kpc) molecular clouds: Orion (combining the Orion A and B clouds), Mon R2, Cep OB3, Taurus, and the {\it c2d} mapped regions of Serpens, Perseus, Ophiuchus, Lupus I, Lupus III, and Lupus IV, and Chamaeleon II.  This study is motivated by the availability of {\it Spitzer} $Space$ $Telescope$  surveys of each of these clouds. With {\it Spitzer}, we are capable of identifying and classifying young stellar objects (YSOs) and, specifically, protostars using mid-infrared photometry.  The protostellar SED peaks longward of the J, H, and K$_{s}$ bands, making it imperative to utilize the longer wavelength photometry to identify and characterize protostars \citep{2008ApJ...674..336G}.  We utilize photometry from the 3.6, 4.5, 5.8, 8.0~$\mu$m bands from the InfraRed Array Camera ({\it IRAC}) and the 24~$\mu$m band on the Multiband Imaging Photometer for {\it Spitzer} ({\it MIPS}) instruments aboard {\it Spitzer} \citep{2004ApJS..154...10F,2004ApJS..154...25R}.  These wavelengths are highly sensitive to the infrared excess exhibited by YSOs, with the excess even more pronounced at wavelengths longer than 24~$\mu$m.  After using the mid-IR colors to identify protostars, we use the slope of the SED and the mid-IR luminosity to estimate the bolometric luminosity of the member protostars and construct luminosity functions for each of the clouds. We then estimate the amount of contamination in our protostar sample due to reddened Class II sources, edge-on disks, and galaxies.

This study expands on previous work in several ways.  The chosen sample of clouds spans a range of environments, from crowded clusters with massive stars to relatively isolated regions of star formation.  We present the first protostellar luminosity functions constructed from the {\it Spitzer} surveys of the Orion, Mon R2 and Cep OB3 clouds, each of which is forming massive stars.  We then compare luminosity functions to test whether the luminosity function is universal, or whether it shows distinct differences between clouds as first suggested in a comparison of the Taurus and Ophiuchus clouds by \cite{1994ApJ...434..614G}.  We also look for spatial variations in the luminosity function within individual clouds by comparing regions of high and low stellar density.  Finally, we compare our luminosity function with recent models of protostellar accretion.

Variations in the protostellar luminosity function are of key interest, since they may trace how differences in the environment affect the star formation process.  The luminosity of a protostar is the sum of the intrinsic luminosity of the central protostar and the accretion luminosity generated by matter falling onto the protostar.  The accretion luminosity  is proportional to the mass accretion rate, and the mass accretion rate is dependent on the rate of mass infall from the envelope \citep[although the accretion rate may not equal the infall rate due to episodic accretion;][]{2005ApJ...633L.137V}.  The mass infall rate, in turn, can depend on the properties of the surrounding gas and the physical mechanism driving infall.  For example, if the mass infall results from the collapse of thermally supported cores, the infall rate increases with increasing sound speed, and hence, increasing temperature \citep{1977ApJ...214..488S}.  On the other hand, if the infall is the result of Bondi-Hoyle accretion from a larger reservoir of gas, the infall rate increases with gas density and stellar mass \citep{2006MNRAS.370..488B}.  Alternatively, if interactions between protostars are important, the luminosity may depend on the degree of clustering.  Thus, variations in the luminosity function can be used to understand the physics that mediates infall and accretion. 
 
Furthermore, variations in the luminosity function may give us some insight into mass segregation.   Higher protostellar luminosities may imply either higher accretion rates - which  can result in higher masses, or higher intrinsic luminosities - which may imply more massive protostars.  Although the relationship between the current luminosity and the ultimate mass is uncertain for protostars, they have the  advantage of being at their birth sites.  In contrast, stars used to determine IMFs have typically dispersed their birth environment and have moved form their birthsites \citep[i.e. Orion, IC 348,][]{2000ApJ...540.1016L}. Thus, protostars provide the means to more directly connect the properties of a forming star and the environment in which it forms.   

In Section 2, we overview the data reduction and photometry. The identification of protostars and the determination of their bolometric luminosities is described in Section 3.  After estimating the contamination in Section 4, the protostellar luminosity functions are presented and analyzed in Section 5.  There we present evidence that the protostellar luminosity functions vary not only between molecular clouds, but also within clouds.  Section 6 compares our luminosity functions with predicted luminosity functions from a variety of accretion models and speculates on the implication of a spatially varying luminosity function for primordial mass segregation.

\section{Data Reduction \& Photometry}

The photometry of the clouds in our sample was assembled from multiple sources, as described in this section.  In addition to using existing source catalogs from the {\it c2d} program, the Taurus Legacy molecular cloud survey, and guaranteed time observations of Taurus, we extracted photometry from {\it Spitzer} surveys of the Orion, Cep OB3, and Mon R2 clouds using PSF fitting photometry techniques. For these three clouds, we applied a new approach for identifying saturated stars and measuring their 24~$\mu$m magnitudes.   We also identified saturated stars in the {\it c2d} clouds that are not in the existing catalogs and measured their photometry.  Finally, we used IRAS 25~$\mu$m photometry to estimate 24~$\mu$m photometry for highly saturated protostars in the Taurus cloud.

We list the IRAC and MIPS photometry for the protostar candidates, identified using the criteria described in Section \ref{sec:criteria}, in Table \ref{table:Table1}.

\subsection{Orion, Cep OB3, and Mon R2 Photometry}
\label{sec:saturation}

Aperture photometry of the IRAC data (Program IDs (PID) 43, 50, 30641, 50070) in Orion were taken from Megeath et al. (in prep).  The IRAC aperture photometry from Cep OB3 and Mon R2 (PID 20403) were taken from Gutermuth et al. (in prep) with no modification.  MIPS data for Orion (PID 43, 47, 58, 30641, 50070), and Cep OB3 and Mon R2 (PID 20403) were reduced, calibrated and mosaicked using the MIPS instrument team’s Data Analysis Tool \citep{2005PASP..117..503G}.  An additional MIPS field in the Cep OB3 cloud containing an extension of the Cep OB3b cluster (PID 40147) was reduced using Cluster Grinder \citep{2009ApJS..184...18G}. Point spread function (PSF) fitting photometry was used for the MIPS 24 $\mu$m sources in each of these regions due the lower angular resolution in the wavelength band, the resulting confusion with other sources, and also nebulosity present at this wavelength. The PSF fitting photometry was performed directly on the 24 $\mu$m data  mosaic. The steps of the PSF extraction are described below.

First, {\it MIPS} 24~$\mu$m sources were identified using the source finding routine in PhotVis \citep{2008ApJ...674..336G}.  The photometry was then extracted by PhotVis using an aperture size of 5 pixels and sky annulus between 12 and 15 pixels, with each pixel having a size of 1."25.  Point-like sources were automatically selected by PhotVis, but careful visual inspection of the images added a few sources per image, mostly in nebulous regions.  A zero-point magnitude of 16.48 was adopted; this included the aperture correction from 5 pixels to infinity of 1.69975.  The images were in units of DN/s.  The sources were detected in PhotVis above a 6 $\sigma$ threshold, then each image was carefully inspected for missing sources.

Next, we created two PSFs using a sample of bright but not saturated stars; these were selected to be relatively isolated and in regions with little nebulosity.  Typically 15 sources were used for a given mosaic, but in a subset of mosaics there were fewer than 15 suitable sources available.  We used two PSFs; one which would be used to extract photometry from the unsaturated sources and one used for the saturated sources, the latter of which was larger in order to encompass the extended PSF wings of the saturated sources.  The larger PSF used for saturated sources has a radius of 35 pixels and a fitting radius of 10 pixels, while the PSF used on unsaturated sources has a radius of 25 pixels and fitting radius of 2 pixels.  

We then fit the appropriate PSF to the sources using the IDL implementation of {\it NSTAR} from the IDL astronomy library \citep{1993ASPC...52..246L}.  To minimize the effect of crowding on the photometry, the sources were first grouped together in clusters of potentially overlapping source PSFs using the routine {\it GROUP}.  This routine groups together sources which were within 8 pixels of one of the other group members; the sources in the group were then fit simultaneously using {\it NSTAR}.  Sources which contained pixels with signals greater than the saturation limit were flagged.  To fit the saturated sources, we modified {\it NSTAR} to ignore saturated pixels, which were flagged by a ``NaN'' value in the image.  Unsaturated sources were fit with the smaller PSF while saturated stars were fit with the PSF with the larger size and fitting radius to ensure there were enough unsaturated pixels to provide a robust fit.  Any unsaturated sources which were grouped together with saturated sources were then treated like saturated sources for PSF fitting and fit using the larger PSF.  We then used {\it NSTAR} to extract fluxes by first extracting fluxes for grouped unsaturated sources and then for grouped saturated sources.  Afterward, the PSFs were scaled  by the photometry and subtracted out with the {\it SUBSTAR} routine; this process was done twice, once first for the unsaturated and once for the saturated sources. The residual image was then inspected for sources which had not been extracted.  These sources, typically one or fewer for each of the mosaics, were added into the photometry.  The process of grouping, fitting, and extraction was iterated until we found no more new sources.

Very saturated sources (those with many saturated pixels and for which PhotVis had difficulty estimating fluxes) may not be easily fit by {\it NSTAR}.  These sources were then given estimated magnitudes and run through {\it NSTAR} using these estimations.  After flux extraction, a residual image was created by {\it SUBSTAR}, and visually inspected for poorly subtracted PSFs.  Saturated sources were typically over-subtracted and were apparent in the residual image.  The input magnitudes initially from PhotVis aperture photometry or from our initial estimation were adjusted until the residual in the image was minimized.  Photometric uncertainties for these sources are due to the varied environments of these sources, including areas of crowding or nebulosity, and were then found by determining the change in magnitude needed to create a noticeable over- or under-subtracted residual.  This constrains the source magnitude to within typically $\pm$0.25~$mag$.  We fit 6 protostar candidates in Orion in this way.  For some sources, positions needed to be adjusted after {\it NSTAR} fitting as well, which was done so that the center of the PSF fell directly over the center of the source.  Since a formal least-square fit and uncertainty could not be obtained, we use the results of this ``fit by eye'' for these sources.  Three protostar candidates in Orion were given positions in this way.  The process of grouping, fitting, and extraction was iterated while we found more new sources.  

Since each {\it MIPS} frame is the subtraction of readouts from the beginning and end of the integration, some very saturated pixels had  values below our adopted threshold for saturation.  These saturated pixels can be identified by comparing the original image with a ``fake" image. We built the ``fake" image by first creating an image of the background nebulosity with the stellar sources subtracted.  This was done by first running {\it SUBSTAR} and smoothing the resulting image to minimize the artifacts from the subtraction, including the reduction of ``holes" in the image due to the over-subtraction of saturated sources.  The subtracted stars were then added back to the smoothed image using one of our two PSFs scaled to the best fit photometry.  Where the resulting simulated data exceeded the saturation limit we masked out pixels.  The entire photometric extraction process was then repeated with the newly identified saturated pixels masked.  This process was repeated until no new saturated pixels were found.

The uncertainties listed in Table \ref{table:Table1} are the relative uncertainties of the data taken into account the photon noise and variations of the signal in the sky annulus.  The 24~$\mu$m uncertainties are those returned by the NSTAR program.  For the highly saturated stars that are ``fit by eye", the uncertainties are given by the change in the magnitude to create a noticeable change in the residual and are noted in Table \ref{table:Table1}.  The reported uncertainties do not include absolute uncertainties due to calibration errors; these uncertainties are 5\% for the IRAC and MIPS instruments.

\subsection{c2d Sources}

Photometry for Serpens, Perseus, Ophiuchus, Chamaeleon, and Lupus were obtained from the Cores to Disks ({\it c2d}) Spitzer Legacy project 4$^{th}$ high-reliability data release\footnotemark[1].\footnotetext[1]{From $http://irsa.ipac.caltech.edu/data/SPITZER/C2D/$}  We used only the fluxes that were not band-filled\footnotemark[2]
\footnotetext[2]{Band-filling is described in the {\it c2d} Final Delivery Document, available at $http://irsa.ipac.caltech.edu/data/SPITZER/docs/spitzermission/observingprograms/legacy/c2d/$}, thus requiring that the source was detected in a specific band before adopting the tabulate photometry.  These sources were identified in the {\it c2d} photometry tables as having an entry of `-2' in the ``image type'' column in the high reliability catalog.  In addition to the high-reliability catalog, we inspected the 24~$\mu$m images of these clouds to search for sources that were not included in the {\it c2d} catalog due to saturation.  For this purpose, we used the 24~$\mu$m image of Serpens from \cite{2007ApJ...669..493W} and {\it c2d} images\footnotemark[1] for the Perseus, Ophiuchus, Lupus, and Chamaeleon II clouds.  By comparing source list positions with the images, we identified sources which do not appear in the  {\it c2d} catalog and appeared saturated in the images.  We then attempted to fit these sources using the PSF technique described in Section \ref{sec:saturation}.  We were successful in obtaining 24~$\mu$m fluxes for 9 saturated sources across all of the {\it c2d} clouds using the modified version of {\it NSTAR}.  We used a sample of unsaturated sources from the Ophiuchus cloud to compare our photometry extraction method with that of the {\it c2d} survey.  We found in a comparison of 72 sources, the mean difference in $m_{24}$ which we extracted and $m_{24}$ from the {\it c2d} data release is 0.0061 $mag$ with a standard deviation of 0.12 $mag$.  Similarly, we compared our extracted Serpens $m_{24}$ with that from the {\it c2d} data release, and found the median difference among 168 sources to be 0.051 $mag$ with a standard deviation of 0.13 $mag$.  These differences are well below the uncertainties in the flux calibration and the uncertainties in the measurements.

The positions and 24~$\mu$m photometry for the saturated sources are listed along with the rest of the photometry in Table \ref{table:Table1}.  Also in Table \ref{table:Table1} are the uncertainties listed for the 2MASS, IRAC, and unsaturated MIPS sources are adopted from the {\it c2d} catalog.

\subsection{Taurus Sources}

We used Taurus photometry from the {\it Spitzer Legacy Taurus Molecular Cloud Survey} data release S14\footnotemark[3].  \footnotetext[3]{From http : //ssc.spitzer.caltech.edu/spitzermission/observingprograms/legacy/taurus/}  We add to our sample 13 sources in the L1551 region from \cite{2009ApJS..184...18G}; L1551 is a small satellite of the Taurus region with several well known protostars which was not covered by the {\it Spitzer Legacy} map of Taurus \citep{1987ApJS...63..645U}.  L1551 contains some of the most luminous known protostars in Taurus and thus we desired inclusion of these sources in our Taurus photometry sample.  For 2 protostars from \cite{2009ApJS..184...18G} in L1551 and 6 from the Legacy survey which do not have 24~$\mu$m detections due to saturation, we used IRAS fluxes at 25~$\mu$m from the Faint Source Catalog except for L1551 IRS5, which is from the IRAS Point Source Catalog.  We then convert from 25~$\mu$m flux to 24~$\mu$m flux using the approximation $F_{24} = K \times F_{25}$ where K = 0.986 for a cool blackbody (T = 70 K) ({\it MIPS} Handbook, version 2.0, 2010).  The sources receiving their 24~$\mu$m photometry in this way are L1551 IRS5, IC 2087 IR, 04365$+$2535, HL Tau, GV Tau, 04239$+$2436, 04278$+$2253AB, and 04361$+$2547AB.

In Table \ref{table:Table1} we list the photometry we have adopted for the Taurus protostar candidates.  For most sources the photometry is from the Taurus Legacy Survey; we note in this table the sources for which we use photometry and adopted uncertainties from \cite{2009ApJS..184...18G}, the IRAS Faint Source Catalog, and the IRAS Point Source Catalog.

Taurus is a relatively well-studied cloud and it is important to ensure that known Taurus protostars are included or accounted for in our sample.  We compare our list with a list of known protostars from \cite{2008ApJS..176..184F}.  We find that all but two of the 28 sources listed in their Table 1 are also in our photometry sample; the remaining sources fall off the Taurus Survey map and L1551 field.  We compare the results of our protostar selection criteria with the protostars tabulated in \cite{2008ApJS..176..184F} in Section \ref{sec:criteria}.

\begin{sidewaystable}\tiny
\caption{Properties of Protostar Candidates}
\begin{tabular}{llllllllllll}
\hline\hline
 &   & \multicolumn{3}{c}{$2MASS$}  &  \multicolumn{4}{c}{{\it IRAC}} & \multicolumn{1}{c}{{\it MIPS}} \\
\hline
RA & Dec & J (unc) & H (unc)  & $K_{s}$ (unc) & 3.6  $\mu$m (unc) & 4.5 $\mu$m (unc)  & 5.8 $\mu$m (unc)  & 8.0 $\mu$m (unc)   & 24 $\mu$m (unc) & $\alpha$ & $log(L/L_{\odot})$ \\
\hline
16:21:45.12 &  -23:42:31.6 &  16.25$\pm$0.11 &  14.72$\pm$0.06 &  13.58$\pm$0.05 &  11.77$\pm$0.05 &  10.74$\pm$0.05 &  10.50$\pm$0.05 &  9.13$\pm$0.05 &  3.88$\pm$0.10 &  0.920 &  -1.29 \\
16:23:05.43 &  -23:02:56.7 &  14.83$\pm$0.03 &  13.86$\pm$0.02 &  13.30$\pm$0.03 &  12.69$\pm$0.05 &  12.07$\pm$0.05 &  11.31$\pm$0.05 &  10.16$\pm$0.05 &  4.76$\pm$0.10 &  1.000 &  -1.43 \\
16:25:27.56 &  -24:36:47.5 &  13.45$\pm$0.02 &  12.80$\pm$0.02 &  12.63$\pm$0.03 &  10.97$\pm$0.05 &  10.02$\pm$0.05 &  8.96$\pm$0.05 &  7.57$\pm$0.05 &  4.81$\pm$0.10 &  0.020 &  -1.36 \\
16:26:17.22 &  -24:23:45.1 &  \nodata &  15.46$\pm$0.11 &  12.25$\pm$0.02 &  9.48$\pm$0.05 &  8.28$\pm$0.05 &  7.13$\pm$0.05 &  6.08$\pm$0.05 &  2.86$\pm$0.10 &  0.190 &  -0.86 \\
16:26:25.46 &  -24:23:01.3 &  \nodata &  15.29$\pm$0.15 &  13.34$\pm$0.07 &  11.19$\pm$0.05 &  10.13$\pm$0.05 &  9.31$\pm$0.05 &  8.34$\pm$0.05 &  3.01$\pm$0.10 &  1.020 &  -0.94 \\
16:26:25.62 &  -24:24:28.8 &  \nodata &  \nodata &  \nodata &  \nodata &  12.26$\pm$0.08 &  14.48$\pm$4.11 &  10.87$\pm$0.07 &  4.11$\pm$0.10 &  2.340 &  -1.01 \\
16:26:40.46 &  -24:27:14.3 &  \nodata &  15.71$\pm$0.11 &  12.34$\pm$0.02 &  9.91$\pm$0.05 &  9.01$\pm$0.05 &  8.20$\pm$0.05 &  7.22$\pm$0.05 &  3.15$\pm$0.10 &  0.330 &  -1.14 \\
16:26:44.19 &  -24:34:48.3 &  16.76$\pm$0.14 &  13.76$\pm$0.04 &  11.61$\pm$0.02 &  7.68$\pm$0.05 &  5.96$\pm$0.06 &  4.63$\pm$0.08 &  3.64$\pm$0.06 &  0.03$\pm$0.01 &  0.580 &  0.335 \\
16:26:48.46 &  -24:28:38.6 &  \nodata &  13.93$\pm$0.04 &  11.21$\pm$0.02 &  8.87$\pm$0.05 &  8.02$\pm$0.05 &  7.38$\pm$0.05 &  6.59$\pm$0.05 &  3.02$\pm$0.10 &  -0.12 &  -0.88 \\
16:26:53.46 &  -24:32:36.1 &  \nodata &  \nodata &  13.12$\pm$0.03 &  10.56$\pm$0.05 &  9.73$\pm$0.05 &  9.29$\pm$0.05 &  8.99$\pm$0.05 &  4.89$\pm$0.10 &  -0.23 &  -1.65 \\
16:27:05.24 &  -24:36:29.5 &  \nodata &  \nodata &  14.45$\pm$0.09 &  11.66$\pm$0.06 &  10.56$\pm$0.06 &  9.60$\pm$0.05 &  8.34$\pm$0.05 &  2.98$\pm$0.10 &  1.270 &  -0.87 \\
16:27:06.75 &  -24:38:14.8 &  \nodata &  14.30$\pm$0.05 &  10.97$\pm$0.02 &  7.67$\pm$0.06 &  6.59$\pm$0.06 &  5.79$\pm$0.05 &  4.91$\pm$0.05 &  1.02$\pm$0.10 &  0.230 &  -0.27 \\
16:27:17.57 &  -24:28:56.2 &  \nodata &  14.42$\pm$0.09 &  11.58$\pm$0.04 &  8.36$\pm$0.05 &  7.35$\pm$0.05 &  6.51$\pm$0.05 &  5.95$\pm$0.05 &  2.41$\pm$0.10 &  -0.12 &  -0.64 \\
16:27:18.36 &  -24:39:14.5 &  \nodata &  15.53$\pm$0.09 &  12.23$\pm$0.02 &  9.52$\pm$0.05 &  8.49$\pm$0.05 &  7.75$\pm$0.05 &  6.94$\pm$0.05 &  3.64$\pm$0.10 &  -0.15 &  -1.09 \\
16:27:21.78 &  -24:29:53.1 &  \nodata &  15.38$\pm$0.09 &  10.83$\pm$0.02 &  6.95$\pm$0.07 &  5.72$\pm$0.06 &  4.76$\pm$0.05 &  3.92$\pm$0.06 &  0.54$\pm$0.10 &  0.070 &  0.057 \\
16:27:24.58 &  -24:41:03.0 &  \nodata &  \nodata &  13.67$\pm$0.05 &  9.29$\pm$0.05 &  7.84$\pm$0.05 &  6.95$\pm$0.05 &  6.28$\pm$0.05 &  1.82$\pm$0.10 &  0.550 &  -0.54 \\
16:27:26.26 &  -24:42:46.0 &  \nodata &  15.18$\pm$0.07 &  12.66$\pm$0.02 &  10.63$\pm$0.05 &  9.86$\pm$0.05 &  9.12$\pm$0.05 &  8.29$\pm$0.05 &  4.52$\pm$0.10 &  0.030 &  -1.55 \\
16:27:27.98 &  -24:39:33.3 &  \nodata &  13.68$\pm$0.08 &  10.38$\pm$0.03 &  6.46$\pm$0.14 &  4.98$\pm$0.09 &  3.98$\pm$0.09 &  3.60$\pm$0.11 &  -0.67$\pm$0.18 &  0.360 &  0.393 \\
16:27:30.90 &  -24:27:33.1 &  \nodata &  \nodata &  14.86$\pm$0.14 &  10.76$\pm$0.05 &  9.12$\pm$0.05 &  7.93$\pm$0.05 &  6.78$\pm$0.05 &  3.75$\pm$0.10 &  0.280 &  -1.20 \\
16:27:37.22 &  -24:42:37.8 &  \nodata &  14.52$\pm$0.05 &  11.46$\pm$0.03 &  8.72$\pm$0.05 &  7.76$\pm$0.05 &  7.13$\pm$0.05 &  6.39$\pm$0.05 &  2.77$\pm$0.10 &  -0.10 &  -0.81 \\
16:27:38.92 &  -24:40:20.4 &  16.54$\pm$0.12 &  13.91$\pm$0.04 &  12.29$\pm$0.02 &  10.62$\pm$0.05 &  9.57$\pm$0.05 &  8.70$\pm$0.05 &  7.70$\pm$0.05 &  2.98$\pm$0.10 &  0.750 &  -0.93 \\
16:27:41.60 &  -24:46:44.6 &  17.20$\pm$0.20 &  15.33$\pm$0.07 &  13.55$\pm$0.04 &  11.32$\pm$0.05 &  10.28$\pm$0.05 &  9.38$\pm$0.05 &  8.38$\pm$0.05 &  5.04$\pm$0.10 &  0.050 &  -1.70 \\
16:27:45.76 &  -24:44:53.8 &  17.38$\pm$0.23 &  14.54$\pm$0.06 &  12.46$\pm$0.03 &  9.70$\pm$0.05 &  8.71$\pm$0.05 &  7.97$\pm$0.05 &  7.27$\pm$0.05 &  3.88$\pm$0.10 &  -0.18 &  -1.18 \\
16:27:48.22 &  -24:42:25.4 &  \nodata &  \nodata &  14.42$\pm$0.07 &  13.41$\pm$0.05 &  12.28$\pm$0.05 &  11.25$\pm$0.06 &  9.14$\pm$0.05 &  3.98$\pm$0.10 &  1.680 &  -1.09 \\
16:28:21.60 &  -24:36:23.4 &  \nodata &  \nodata &  \nodata &  14.20$\pm$0.07 &  12.12$\pm$0.06 &  11.02$\pm$0.06 &  10.78$\pm$0.05 &  4.75$\pm$0.10 &  1.390 &  -1.56 \\
16:28:57.85 &  -24:40:54.8 &  \nodata &  15.79$\pm$0.16 &  13.87$\pm$0.05 &  11.15$\pm$0.05 &  10.06$\pm$0.05 &  9.13$\pm$0.05 &  8.06$\pm$0.05 &  4.12$\pm$0.10 &  0.430 &  -1.45 \\
16:31:35.64 &  -24:01:29.3 &  15.51$\pm$0.06 &  11.83$\pm$0.02 &  9.22$\pm$0.02 &  6.70$\pm$0.06 &  5.68$\pm$0.06 &  4.91$\pm$0.05 &  4.11$\pm$0.06 &  0.87$\pm$0.10 &  -0.17 &  0.043 \\
16:31:36.77 &  -24:04:19.7 &  15.74$\pm$0.07 &  13.85$\pm$0.06 &  12.50$\pm$0.03 &  11.46$\pm$0.05 &  11.05$\pm$0.05 &  11.10$\pm$0.05 &  11.14$\pm$0.05 &  3.96$\pm$0.10 &  0.810 &  -1.36 \\
16:31:52.45 &  -24:55:36.1 &  \nodata &  14.84$\pm$0.06 &  12.72$\pm$0.02 &  10.33$\pm$0.05 &  9.12$\pm$0.05 &  8.06$\pm$0.05 &  6.87$\pm$0.05 &  2.28$\pm$0.10 &  0.920 &  -0.61 \\
\label{table:Table1}
\end{tabular}
\begin{tablenotes}
\item[*]{Table 1 in its entirety is available online.  A portion is shown here for guidance regarding its form and content.}
\end{tablenotes}
\end{sidewaystable}

\section{Identifying Protostars and Measuring Their Luminosity} 

Here we describe the methodology for selecting protostar candidates and estimating their bolometric luminosities.  We give the criteria used to identify protostar candidates using color-magnitude diagrams in Section \ref{sec:criteria}.  We then discuss the spectral energy distributions (SEDs) and the technique used to determine bolometric luminosity in Section \ref{sec:bol_lum}. This technique is established empirically using the SED slopes and mid-infrared luminosity of protostars in the {\it c2d} catalogs with known bolometric luminosities.  We then create luminosity functions for each of our surveyed clouds (Section \ref{sec:PLFs}).

\subsection{Protostar Candidate Selection}
\label{sec:criteria}

The {\it Spitzer} space telescope, with its ability to rapidly map entire molecular clouds with $\le$ 5'' resolution and high sensitivity between 3.6 and 24~$\mu$m, has dramatically increased the number of known protostars within 1 kpc \citep[e.g.,][]{2009ApJ...692..973E,2009ApJS..181..321E,2009ApJS..184...18G}. Protostars can be readily readily identified by their flat or rising spectral energy distributions over the {\it Spitzer} wavelength range \citep{1994ApJ...434..614G}.  Several schemes for identifying protostars in {\it Spitzer} surveys have been developed; these rely either on fitting models to the SED (ranging from simple power-laws  to detailed radiative models) or on measuring mid-IR colors that depend on the SED slope \citep{2004ApJS..154..363A,2004ApJS..154..367M,2004ApJS..154..379M,2004ApJ...617.1177W,2006ApJS..167..256R,2007ApJ...663.1149H,2007ApJ...669..493W,2009ApJS..184...18G}.  In this paper, we identify protostars using criteria first developed by \cite{2009AJ....137.4072M} and Megeath et al. in prep. for protostars with 24~$\mu$m detections.  These criteria are satisfied by SEDs which have a power-law slope (defined as $\alpha$ = $d log(\lambda$$F_{\lambda})/d log({\lambda})$) between the 4.5~$\mu$m and 24~$\mu$m bands greater than $\alpha >$ -0.3.  This corresponds to the lowest $\alpha$ for flat spectrum sources in \cite{1994ApJ...434..614G}. We sought sources where the emission at 24~$\mu$m was dominated by thermal emission from infalling envelopes \citep{2003ApJ...598.1079W}.  These include flat spectrum sources, Class I sources and Class 0 sources \citep{1984ApJ...287..610L,1994ApJ...434..330C,2007ApJ...669..493W,2009ApJ...692..973E}.

Before applying the criteria designed to identify protostar candidates, several detection and signal-to-noise criteria were applied to the point sources recovered by our photometry. We began by restricting ourselves to the set of protostars with detection in the {\it MIPS} 24~$\mu$m band, which detects thermal emission from the envelope.  In contrast, the shorter wavelength IRAC bands are typically dominated by light from the inner disk which is scattered by dust in the envelope \citep{2003ApJ...598.1079W}. In the next section, we will show that the 24~$\mu$m band is needed to estimate luminosities of the protostars; thus likely protostars without 24~$\mu$m detections are ignored.  Protostars which are located in saturated areas of our 24~$\mu$m maps, i.e.the Orion Nebula or NGC~2024, were not included in this survey.  To minimize contamination from galaxies, we required that protostar candidates be projected on regions of the clouds where $A_{V}$ $>$ 3, as determined from extinction maps of the clouds (these maps are discussed at the end of this section). The contamination from galaxies was further reduced by adopting a maximum 24~$\mu$m magnitude \citep{2009AJ....137.4072M}; as discussed later, a different limit was set for each cloud.  Finally, in each set of criteria, we required that the utilized photometry had uncertainties in J, H, K$_{s}$, 3.6, and 4.5~$\mu$m $<$ 0.1 $mag$ and uncertainties in 5.8 and 8.0~$\mu$m $<$ 0.15 $mag$, and a detection at 24~$\mu$m.

\begin{figure}
\centering
\includegraphics[width=0.8\textwidth]{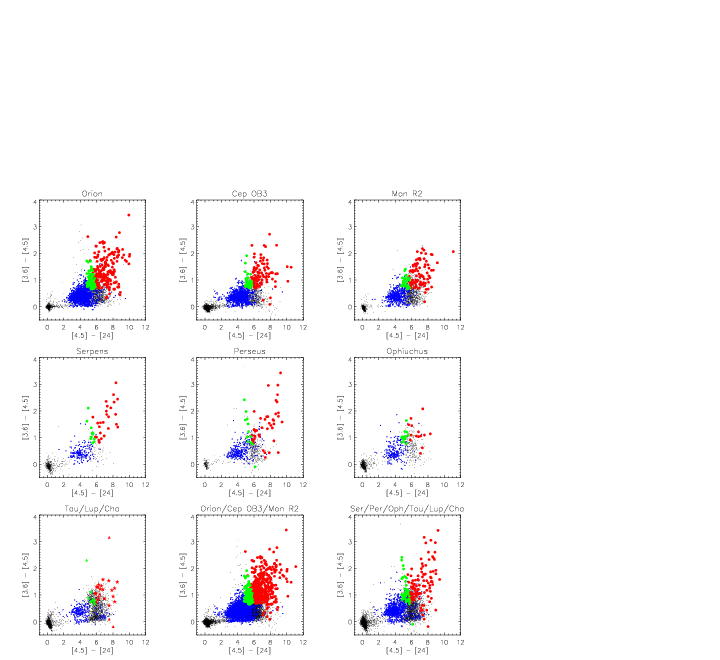}
\caption{Color-color diagrams for {\it Spitzer} identified protostars.  The rising spectrum protostars are shown in red and flat spectrum protostars are in green.  The sources we identify as Class IIs are shown in blue.  The remaining sources, shown as black dots, are stars without disks, AGN, and star forming galaxies.  In the panel showing combined Taurus, Lupus, and Chamaeleon cloud data, Taurus protostars are shown as stars, Lupus protostars are shown as triangles, and Chamaeleon protostars are shown as squares. }
\label{fig:fig1}
\end{figure}

The adopted protostar selection criteria were first described in \cite{2009AJ....137.4072M} with a few additional modifications taken from Megeath et al. (in prep).  These selection criteria are initially based on 4.5 and 24~$\mu$m photometry alone since these are typically the most sensitive bands for detecting protostars; at 5.8 and 8.0~$\mu$m the IRAC detector is less sensitive and near 8~$\mu$m the protostellar SED typically dips \citep[Megeath et al., in prep.]{2003ApJ...598.1079W,2009AJ....137.4072M}.  The most deeply embedded sources are sometimes not detected at 3.6~$\mu$m; in these cases, we assigned a photometric upper limit magnitude at 3.6~$\mu$m of 15.5 $mag$ (a conservative limit equal to the $\sim$ 80$\%$ completeness level determined from \cite{2006ApJ...644..307H}; a similar limit is used by Megeath et al. in prep.), to determine lower limits for the colors for these sources.  For sources with uncertainty in $m_{4.5}$ $<$ 0.1 $mag$ and with 24~$\mu$m detections, we required protostars to fulfill the following  criteria:

\begin{eqnarray}
&  [4.5] - [24] \ge 4.761 \\ \nonumber
& \mbox{ and } \\ \nonumber
& [3.6] - [4.5] \ge 0.752 + \sigma_{3.6,4.5}\\ \nonumber
& \mbox{ or } \\ \nonumber
& [4.5] - [24] \ge 5.303 \\ \nonumber
& \mbox{ and } \\ \nonumber
& [3.6] - [4.5] \ge 0.652 + \sigma_{3.6,4.5} \\
\end{eqnarray}

\noindent where for any two photometric bands a and b, $\sigma_{a,b}$ = $\sqrt{\sigma^{2}_{a} + \sigma^{2}_{b}}$.  The colors of $[3.6]-[4.5] = 0.6520$ and $[4.5]-[24] = 4.761$ corresponds to a power-law SED with of $\alpha = -0.3$.  Note that we require that a minimum color of $[4.5]-[24] = 5.303$, corresponding to $\alpha = 0$ for sources where $[3.6]-[4.5] = 0.652$.  This was done to reduce contamination from reddened Class $II$ (i.e. pre-main sequence stars with disks) based on the examination of the color-color diagrams in Figure \ref{fig:fig1} as well as those of regions with $A_{V} < 3$ where few protostars are expected.  

Examination of the data led to the identification of protostellar objects with highly reddened [4.5]-[24] colors but relatively blue [3.6]-[4.5].  These sources showed resolved yet compact scattered light nebulae in $I-band$ images (J. Bally, P. Com); indicating that they were protostars where the blue colors where due to a strong contribution from scattered light.  Based on the colors of these sources, we established the following criteria to ensure that such sources were included in our protostar catalog.  For sources with detections at 4.5, 5.8 and 8~$\mu$m and uncertainties of $\sigma_{4.5} \le 0.1$~$mag$,  $\sigma_{5.8} \le 0.15$~$mag$ and $\sigma_{8.0} \le 0.15$~$mag$, we applied the following criteria:

\begin{eqnarray}
& [4.5] - [24] \ge 7 \\ \nonumber
&  \mbox{ and } \\ \nonumber
& [5.8] - [8.0] \le 1.75
\end{eqnarray}

\noindent where the [5.8] - [8.0] criterion is used to eliminate star forming galaxies with strong PAH emission in the 8~$\mu$m band.  

Sources without a 4.5~$\mu$m detection may still be protostars \citep[see for example][]{2010A&A...518L.122F}. To identify such sources, we used detections in either the 5.8 or 8.0~$\mu$m band.  Sources with detections at the  5.8~$\mu$m band with $\sigma_{5.8}$ $<$ 0.15 were identified as protostar candidates if they satisfied the color criteria

\begin{eqnarray}
& [5.8] - [24] \ge 4.117 \\ \nonumber
&  \mbox{ and } \\ \nonumber
& [3.6] - [5.8] \ge 1.296 + \sigma_{3.6,4.5}
\end{eqnarray}

\noindent while for sources with 8~$\mu$m detections and $\sigma_{8.0}$ $<$ 0.15 , we identified protostar candidates with the criteria:

\begin{eqnarray}
& [8.0] - [24] \ge 3.20  \\ \nonumber
&  \mbox{ and } \\ \nonumber
& [3.6] - [8.0] \ge 2.12 + \sigma_{3.6,4.5} 
\end{eqnarray}

\noindent
In both cases, the the colors corresponded to those of a power-law SED of faint protostar candidates where $\alpha = -0.3$. Additionally, we identified sources with less infrared excess and/or fainter $m_{24}$ that are likely more evolved YSOs without envelopes using the criteria for Class II sources from \cite{2009AJ....137.4072M}. Color-color and color-magnitude diagrams for protostar candidates and identified Class II sources are shown in Figures \ref{fig:fig1} and \ref{fig:fig3}, respectively.  We refer to these objects as protostar candidates because we will find that a fraction of these sources may be galaxies, reddened Class II objects, and edge-on disks.  For each protostar candidate, we calculated the spectral slope, $\alpha$, by using a best-fit line to the plot of $log(\lambda$$F_{\lambda})$ vs. $log(\lambda)$ over {\it IRAC} and {\it MIPS} 24~$\mu$m detections.  In Figures \ref{fig:fig1} and \ref{fig:fig3}, we distinguish the flat spectrum $-0.3 \ge \alpha \le 0.3$ and rising spectrum $0.3 \ge \alpha$ protostars.  Class II sources are also shown; theses were identified using the criteria described in \cite{2009ApJS..184...18G}.  

\begin{figure}
\centering
\includegraphics[width=0.8\textwidth]{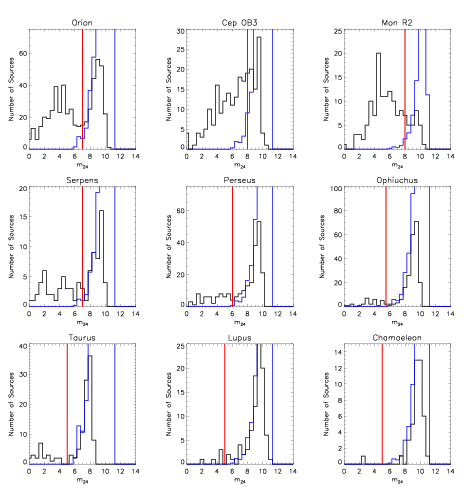}
\caption{Histograms of 24 $\mu$m magnitudes for all sources fitting the protostar criteria and with $A_{V}$ $>$ 3 (black).  Also shown are the histograms of the galaxy contamination estimated from the SWIRE data (blue).  The $m_{24}$ cutoff is shown in red.}
\label{fig:fig2}
\end{figure}

\begin{figure}
\centering
\includegraphics[width=0.8\textwidth]{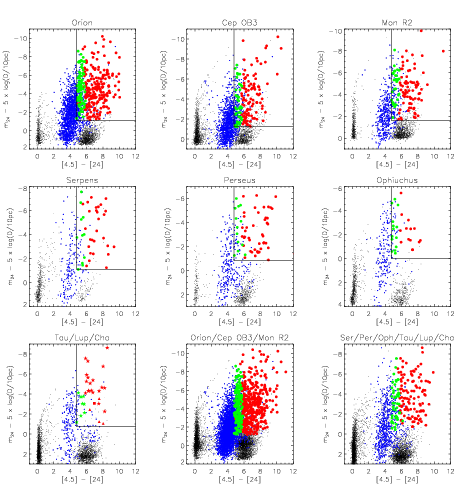}
\caption{Color-magnitude diagrams for {\it Spitzer} identified protostars.  The solid lines show the $m_{24}$ cutoff and the [4.5] - [24] color cutoff.  The 24 $\mu$m magnitude is corrected for distance but not reddening.  Galaxies comprise a distinct clump of fainter sources (at 24 $\mu$m) with colors similar to those of protostars, falling near [4.5] - [24] = 6.  Symbols and colors are the same as Figure \ref{fig:fig1}.}
\label{fig:fig3}
\end{figure}

A  24~$\mu$m  magnitude cutoff was imposed to minimize background galaxy contamination \citep{2009AJ....137.4072M}.  At magnitudes fainter than this cutoff, the contamination rises quickly and surpasses the number of protostar candidates.  To determine the optimal value of the cutoff, a histogram of the 24~$\mu$m magnitudes for sources satisfying the above color criteria was created for each cloud (Figure \ref{fig:fig2}).  Galaxy contamination was estimated by applying our protostar candidate identification criteria to the SWIRE\footnotemark[4]\footnotetext[4]{We used the 4.2~$deg^2$ Elais-N2 field centered at l = 65.022841, b = 42.163632 from http://swire.ipac.caltech.edu/swire/swire.html} sample assuming the density of galaxies in the SWIRE field is similar to that toward our sample of molecular clouds.  Our sample of molecular clouds are distributed over a broad range in Galactic coordinates from $\sim$30$^\circ$ $<$ l $<$ $\sim$350$^\circ$ and $\sim$-22$^\circ$ $<$ b $<$ $\sim$20$^\circ$ .  Since the $m_{24}$ cutoff is primarily concerned with extragalactic contamination, which are distributed approximately evenly over the sky, the SWIRE field is appropriate for estimating contamination due to galaxies.  Extragalactic wide field surveys such as SWIRE have been used by a number of investigators to estimate galactic contamination \citep{2006ApJ...644..307H,2008ApJ...674..336G}.  This number of contaminants was scaled to the size of the $A_{V}$ $>$ 3 areas of the coverage maps for each region, namely 4.25, 8.88, 0.83, 0.45, 2.46, 0.86, 5.49, 1.37, and 0.32~deg$^{2}$ for Ophiuchus, Taurus, Lupus, Chamaeleon II, Perseus, Serpens, Orion, Cep OB3, and Mon R2, respectively.  These are plotted on the same histogram as the candidate protostars toward the clouds (Figure \ref{fig:fig2}). The adopted $m_{24}$ cutoff is shown for each region.  In the following analysis, we only include protostars that are brighter than the adopted cutoff magnitude.  

For the {\it c2d} clouds, we compared our sample of protostars with the sample of 132 c2d sources with envelopes detected at 1.3 mm \citep{2009ApJ...692..973E}.   We classified 88 of these sources as protostars and 14 as Class II sources.  Of the remaining 30 {\it c2d} envelope sources, we rejected 17 as protostars because of missing or band-filled 24~$\mu$m detections or because $m_{24}$ exceeded the cutoff magnitude, four are rejected for uncertainties above our requirements for protostar selection, three satisfy our criteria but are rejected because they fall off the available $A_{V}$ maps or do not have an $A_{V}$ $>$ 3, three sources are rejected because they are not consistent with our criteria, and three sources are found only in the full {\it c2d} source catalog and not found in the {\it c2d} high reliability catalog.  Given the relatively low angular resolution of the Bolocam data ($30 ''$), we suggest that some of the envelope sources with Class II colors may be due to chance coincidences between pre-main sequence stars and dense cloud material.  In other cases, the 1.3 mm measurements may be detecting massive disks.  Additionally, we identified 52 protostars not identified as likely envelopes sources since they do not have 1.3 mm detections of envelopes; we expect these sources to have envelope mass less than 0.5~M$_{\odot}$ \citep{2009ApJS..181..321E}.  

In the Taurus region, we compared the candidate protostars identified using our criteria with the protostars studied in \cite{2008ApJS..176..184F}.  We find that of the 26 protostars in \cite{2008ApJS..176..184F}, 18 sources also satisfy our criteria, although one source has $A_{V}$ $<$ 3.  Of the remaining sources, six of the \cite{2008ApJS..176..184F} protostars we identify as Class II sources, one has a 24~$\mu$m magnitude below or flux cutoff and one has does not have $24~\mu m$ detections. We note that \cite{2008ApJS..176..184F} also find that several of their objects may be sources transitioning between the protostellar and Class II phase. 

\begin{figure}
\centering
\includegraphics[width=0.8\textwidth]{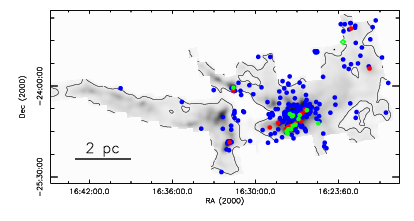}
\caption{(a). The Ophiuchus $A_{V}$ map.  Shown are all Class IIs (blue), rising spectrum protostar candidates (red), and flat spectrum (green) protostar candidates with $A_{V}$ $>$ 3 (circles) and with $A_{V}$ $<$ 3 (diamonds), as well as the $A_{V}$ = 3 contours.}
\label{fig:fig4}
\figurenum{4a}
\end{figure}

Figure \ref{fig:fig4} shows the extinction maps and spatial distributions of all protostar candidates found in our study using our selection criteria as well as the sources that fit the selection criteria but reside in $A_{V}$ $<$ 3 regions, again separating the flat and rising SED sources.  We took the $A_{V}$ maps for the {\it c2d} clouds from the {\it c2d} Legacy project 4$^{th}$ high-reliability data release\footnotemark[5]\footnotetext[5]{From http://irsa.ipac.caltech.edu/data/SPITZER/C2D/.}, the map of Orion is from Megeath et al. (in prep.), the maps of Cep OB3 and Mon R2 clouds are from Gutermuth et al. (in prep.), and the Taurus map is from \cite{2010A&A...512A..67L}.  Also shown in this figure is an $A_{V}$ $>$ 3 contour on each of the maps.

\subsection{Determination of Bolometric Luminosity}
\label{sec:bol_lum}

The majority of protostellar luminosity is emitted at wavelengths longer than 24~$\mu$m, wavelengths at which data are not available for most sources in our study.  Since the total bolometric luminosity, $L_{bol}$, is dominated by the flux at these longer wavelengths,  we need a method for estimating $L_{bol}$ using the luminosities integrated over the available wavelength bands.  We present here a method for using the SED slope and the total mid-infrared luminosity in the {\it Spitzer} 3.6- 24~$\mu$m bands, $L_{MIR}$, to estimate $L_{bol}$.  We first discuss the calculation of the SED slope and of $L_{MIR}$ and then the technique used to estimate bolometric luminosity.
  
We created SEDs for each of our protostars and used the {\it Spitzer} 3.6 - 24~$\mu$m bands to determine the SED slope.  We calculated $\alpha$ by using a best-fit line to the plot of log($\lambda F_{\lambda}$) vs.~log($\lambda$) over {\it IRAC} and {\it MIPS} 24~$\mu$m detections.  Mid-IR luminosities were calculated by integrating the SED over the available fluxes from J, H, K$_{s}$, and {\it IRAC} bands.  Detections were converted to fluxes $F_{\lambda}$, using zero point fluxes of 1594, 1024, 666.8~Jy for J, H, and $K_{s}$ \citep{2003AJ....126.1090C} and 280.9, 179.7, 115.0, 64.13 (IRAC Instrument Handbook, version 1.0), and 7.17~Jy ({\it MIPS} Instrument Handbook, version 2.0) for $F_{3.6}$, $F_{4.5}$, $F_{5.8}$, $F_{8.0}$, and $F_{24}$, respectively.  We divide $F_{24}$ by a color correction of 0.967; for -1 $<$ $\alpha$ $<$ 2 the color correction ranges from 0.960 to 0.967 \citep{2007PASP..119.1038S}.  The adopted {\it IRAC} fluxes are for a flat spectrum source, the color correction ranges from 1 to 1.03 for sources with $\alpha$ = -1 to $\alpha$ = 2 ({\it IRAC} Instrument Handbook, version 1.0).  We use the central wavelength for the J, H, and K$_{s}$ bands and approximate bandwidths from \cite{2003AJ....126.1090C}.  We estimated bandwidths to be 1.073 - 1.397 $\mu$m for $F_{J}$, 1.411 - 1.913 $\mu$m for $F_{H}$, 1.897 - 2.420~$\mu$m for $F_{K_s}$.  We used estimated bandwidths from the {\it IRAC} and {\it MIPS} handbooks 3.175 - 3.925 $\mu$m for $F_{3.6}$, 3.986 - 5.000 $\mu$m for $F_{4.5}$, 5.019 - 6.443 $\mu$m for $F_{5.8}$, 6.420 - 9.324 $\mu$m for $F_{8.0}$, and 20.800 - 26.100 $\mu$m for $F_{24}$.   We used rectangular integration over each band by summing the product of the flux and bandwidth from each band and converting to luminosity using the following equation:

\begin{eqnarray} 
L_{MIR} &=& [19.79 F_{\lambda}(J) + 16.96 F_{\lambda}(H) + 10.49 F_{\lambda}(K_s) + 5.50 F_{\lambda}(3.6) + 4.68 F_{\lambda}(4.5) \nonumber \\
 & & + 4.01 F_{\lambda}(5.8) + 4.31 F_{\lambda}(8.0) + 0.81 F_{\lambda}(24)] \times 10^{-6} \times d^{2} [pc] L_{\odot}
\label{eqn:Lmir}
\end{eqnarray}

\noindent where $d$ is the distance to the cloud in $pc$ and fluxes $F_{\lambda}$ are in $Jy$.  We list the distances in Table \ref{table:sfr_prop}.

\cite{2008ApJS..179..249D} compared the protostellar flux, $\nu$$F_{\nu}$, to the internal luminosity, $L_{int}$, for a grid of radiative transfer models and protostars with complete SEDs from the {\it c2d} survey.  For the 70~$\mu$m band, they found a linear correspondence between $\nu$$F_{\nu}$ and $L_{bol}$.  An approximately linear relationship was also evident at shorter wavelengths, particularly at 24 $\mu$m, but the scatter was much higher, with an order of magnitude variation in luminosity for a given 24~$\mu$m flux.  Because  70~$\mu$m photometry is unavailable for the majority of our sources, we attempt here to reduce the scatter between the mid-IR fluxes and $L_{bol}$.  Since the conversion factor between the mid-IR luminosity and $L_{bol}$ must depend on how rapidly the SED rises at wavelengths longer than 24~$\mu$m, we examined the dependence of the ratio of $L_{MIR}$/$L_{bol}$ (which will be constant if there is a linear dependence between $L_{MIR}$ and $L_{bol}$)  and the slope of the SED between 3.6 and 24~$\mu$m.

To establish a relationship between the slope and $L_{MIR}$/$L_{bol}$, we used protostars selected from the {\it c2d} program with well-established $L_{bol}$.  This sample of {\it c2d} identified protostars spans the range of colors and magnitudes characteristic of protostars as shown in Figures \ref{fig:fig1} and \ref{fig:fig3}.   \cite{2009ApJS..181..321E} determined bolometric luminosities for YSOs in the {\it c2d} catalog using available photometry between 0.36~$\mu$m and 1300~$\mu$m and the method described in \cite{2008ApJS..179..249D}.  We scaled these luminosities to the distances adopted in our study, listed in Table \ref{table:sfr_prop}.  We used 87 YSOs from Serpens, Perseus, Chamaeleon II, Lupus, and Ophiuchus, which are identified as protostar candidates using our criteria and had $L_{bol}$ flagged as good in \cite{2009ApJS..181..321E}, and 70~$\mu$m detections.  We then used the photometry from this sample to determine the slope of the SED and $L_{MIR}$/$L_{bol}$.  

 We found a relationship between the SED slope and ratio of $L_{MIR}$/$L_{{\it c2d}}$ using 66 rising spectrum {\it c2d} protostars.  We choose the slope as a parameter, calculated from 3.6 to 24~$\mu$m as it shows a stronger trend than the slope calculated using only 4.5 and 24~$\mu$m, which are preferentially used in determining protostar candidate status, or the entire range of available wavelengths for all clouds (J-band through 24~$\mu$m).  The 24~$\mu$m flux is useful because it is less affected by inclination effects or the geometry of the outflow cavities, but again we find that using $L_{MIR}$ gives a better fit than using the 24~$\mu$m flux alone.  We found the smallest residuals using a linear relationship between $\sqrt{\frac{L_{MIR}}{L_{{\it c2d}}}}$ and $log(\alpha)$.  We also attempted a polynomial fit to this relationship, but again find that the linear fit has smallest residuals.  Thus, we found that $\sqrt{\frac{L_{MIR}}{L_{{\it c2d}}}}$ goes linearly with $log(\alpha)$.  We expect some intrinsic scatter in the relationship, which is not due to the uncertainty in measurements.  Figure \ref{fig:fig5} shows the best fit for this relationship, and a plot relating $log(\alpha)$ and $L_{{\it c2d}}/L_{MIR}$ luminosity fraction. We found this relationship is

\begin{eqnarray}
& \frac{L_{MIR}}{L_{bol}} = (-0.466\pm0.014 \times log (\alpha)  + 0.337\pm 0.053)^2.
\label{eqn:relationship}
\end{eqnarray} 

\noindent The flat spectrum sources did not follow the same trend. We chose not to fit the flat spectrum sources (-0.3 $<$ $\alpha$ $<$ 0.3) because there were so few in our sample; instead we assumed that flat spectrum sources exhibit a constant ratio in bolometric to mid-IR luminosity.  For these sources we adopted the ratio of $L_{MIR}/L_{bol}$ given in Equation \ref{eqn:relationship} for $\alpha$ = 0.3:

\begin {eqnarray}
& \frac{L_{MIR}}{L_{bol}} = 0.338.
\label{eqn:flat}
\end{eqnarray}

\noindent We find that the mean value of $L_{MIR}$/$L_{{\it c2d}}$ for the 21 flat spectrum sources not used to create the fit is 0.297 with a standard deviation of 0.11, which is consistent with the adopted ratio of Equation \ref{eqn:flat}.

We note that in addition to the internal heating of protostars, the contribution from external heating, $L_{ext}$ may also affect the measured luminosity.   \cite{2001ApJ...557..193E} estimate the effect of $L_{ext}$ is typically of order of 0.1~L$_{\odot}$, which is a small contribution for all but the faintest sources. Therefore, we neglected the effect of $L_{ext}$ on $L_{bol}$.

In Figure \ref{fig:fig6} we compare the luminosities derived using Equations \ref{eqn:relationship} and \ref{eqn:flat} with the bolometric luminosities from \cite{2009ApJS..181..321E}. The protostars from \cite{2009ApJS..181..321E} used to determine Equation \ref{eqn:relationship} range in bolometric luminosity from 0.03 L$_{\odot}$ $\le$ $L_{{\it c2d}}$ $\le$ 19.6 L$_{\odot}$.  The left panel of Figure \ref{fig:fig6} compares the luminosity function constructed from our estimated luminosities to that constructed from the bolometric luminosities in \cite{2009ApJS..181..321E}.  The luminosity functions are similar; a Kolmogorov-Smirnov (K-S) test comparing the two distribution results in a probability of 0.83 that they are from the same parent distribution.  In the right panel we plot the estimated luminosities vs. the luminosities for these sources from \cite{2009ApJS..181..321E}.  In Figure \ref{fig:fig7} we show a histogram of the difference between our estimated $log(L_{bol}/L_{MIR})$  and $log(L_{bol}/L_{MIR})$ derived with the Evans et al. luminosities.  The standard deviation of the difference is 0.35.  From this analysis, we conclude that we can recover the luminosity function of the protostar candidates with reasonable fidelity.    

We can test whether the protostar candidates we identify in Taurus follow a similar relationship.  We compared the luminosities determined with model fits by \cite{2008ApJS..176..184F} with the luminosities determined with our relationship.  These sources are shown in Figure \ref{fig:fig5}.  We find that the Taurus protostar candidates with well-determined bolometric luminosities agree well with our fit; the K-S probability that our Taurus luminosity function is from the same parent distribution as the luminosity function using the model fit luminosities of Table 1 in \cite{2008ApJS..176..184F} is 0.93.  The assumption we will make for the remainder of the paper is that the protostar candidates in more distant clouds forming massive stars will show the same relationship.

Values of $\alpha$ and $L_{bol}$ for each protostar candidate are listed in Table \ref{table:Table1}.

\subsection{Protostellar Luminosity Functions}
\label{sec:PLFs}
We have used the method above to calculate luminosities for each of the protostars in the nine clouds.  In Figure \ref{fig:fig8} we show the resulting luminosity functions.  The protostars of Taurus, Lupus and Chameleon are again combined because of the similar distances, the dispersed star formation of these clouds, and the low number of protostar candidates in these regions.  The individual luminosity functions of these regions and the combined luminosity function are also displayed in Figure \ref{fig:fig9}.  We also show the combined luminosity function for the clouds forming high mass stars (Orion, Cep~OB3 and Mon~R2) and for the clouds forming low to intermediate mass stars (Perseus, Ophiuchus, Taurus, Lupus and Chameleon). 

\section{Contamination}

While our selection technique is designed to minimize the contamination in our final protostar sample, there remain possible sources of contamination.  The most likely contaminants are residual background galaxies, edge-on disk sources, and highly extinguished Class IIs.  A pre-main sequence star with a prominent disk observed at an edge-on inclination can show colors similar to those of a protostar \citep{2008A&A...486..245C}.  Additionally, it has been shown that highly reddened disks in Ophiuchus may look like protostars \citep{2009ApJS..181..321E,2010ApJS..188...75M}.

In this section, the extent of contamination for each cloud is estimated in turn for each type of contamination.  The number of protostellar candidates is listed for each region in Table \ref{table:sfr_prop} both before and after contamination removal (in parenthesis), and the number of protostars or protostar candidates with flat and rising spectra are given. 

\subsection{Extragalactic contamination}
\label{sec:gal_cont}
Galaxies can have colors very similar to those of YSOs \citep{2006ApJ...644..307H,2009ApJS..184...18G}.  Although we have minimized contamination from galaxies by applying the $m_{24}$ cutoff to each of our clouds (Section \ref{sec:criteria}), a small number of extragalactic sources brighter than this limit is expected.   To quantify the contamination from the remaining galaxies on our luminosity functions, we took a sample of SWIRE galaxies from the Elias N2 region scaled to the angular coverage of each cloud, and subjected them to our protostar selection criteria including the $m_{24}$ cutoff.  Any of these known galaxies which were identified as protostars are considered to be contamination.  Using Equations \ref{eqn:relationship} and \ref{eqn:flat}, we determined faux bolometric luminosities for the galaxy contaminants (i.e. the luminosities they would have if they were protostars in the observed cloud) and created luminosity functions for the contaminants.  These are shown in Figure \ref{fig:fig8}.  Although present, galaxies are the smallest source of contamination, comprising only 2.5\% of our protostar candidate sample.

\subsection{Edge-on and Nearly Edge-on Disk Sources}
Pre-main sequence stars seen through their flared disk may have a rising SED and can be mis-identified as a protostar candidates using our criteria.  The fraction of Class IIs seen through their disk is difficult to estimate theoretically, since it depends on poorly constrained properties of the disks, including the amount of flaring and the outer radius of the disk.  Instead, we employed an empirical estimate using the technique of \cite{2009ApJS..184...18G}. We first identified a large cluster of young stars in a region where the gas has been dispersed and the extinction is low.  In our survey, the best example is the Cep OB3b cluster (Allen et al. in prep).  Although the gas has been cleared by the OB stars in the cluster, objects with protostellar-like colors were detected in this low extinction cavity.  Following  \cite{2009ApJS..184...18G}, we assumed that YSOs with protostellar-like colors in the low-extinction regions were edge-on or nearly edge-on disks (hereafter we use `edge-on disks' to refer to disks that are close enough to an edge-on inclination that they are observed through their disks).  We then calculated the ratio of protostars to Class II objects and multiplied this ratio by the number of Class II objects in each cloud to calculate the number of expected edge-on sources.  Since some of the sources in Cep OB3b may be actual protostars which have survived gas dispersal, this assumption gave us an upper limit to the number of edge-on disks sources which have colors similar to protostar candidates.  In the following analysis, we will set the number of edge-on disks equal to this number. 

We identified a region within the Cep OB3b cluster where the total extinction is $A_{V}$ $<$ 3; this low extinction region contained 34 protostar candidates and 568 Class II sources and is shown in the map of Cep OB3 in Figure \ref{fig:fig4}.  The number of Class IIs, however, was for the entire range of magnitudes and did not include a cutoff in $m_{24}$.  Furthermore, these sources were not corrected for possible extragalactic contamination.  To calculate the appropriate ratio, R, which imposed a cutoff at $m_{24}$, we calculated the ratio of edge-on disks to Class II objects using the following equation:

\begin{eqnarray}
& R = \frac{N_{Cep OB3b}^{p}-N_{gal}^{p}}{N_{Cep OB3b}^{d}-N_{gal}^{d}} \nonumber \\
\end{eqnarray}

\noindent where $N_{Cep OB3b}^{p}$ is the number of sources with protostellar-like colors in Cep OB3b with $m_{24}$ $<$ cutoff, $N_{Cep OB3b}^{d}$ is the number of Class II sources with $m_{24}$ brighter than the cutoff, and $N_{gal}^{p}$ and $N_{gal}^{d}$ are the expected contamination from galaxies with protostellar-like and disk-like colors, respectively.  Although R is an upper limit, we treat $R$ as the actual fraction of edge-on disks in the remainder of this paper and thus we may overestimate the contamination by edge-on disks.  The number of edge-on disks in the $A_{V}$ $>$ 3 region was calculated as

\begin{eqnarray}
& N_{edge} = R \times [N_{p}+N_{CII}-N_{gal}^{d}-N_{gal}^{p}] \nonumber \\
\end{eqnarray}

\noindent where R is calculated for Cep OB3b using the $m_{24}$ cutoff of that cloud; but $N_{CII}$ and $N_{p}$ are the number of Class II sources and protostars brighter than the $m_{24}$ cutoff determined for the $A_{V}$ $>$ 3 region of the cloud.  We included the number of protostars ($N_{p}$) since a fraction of our protostars may have been edge-on disks.  This overestimates slightly the number of edge-on disks; however, not including $N_{p}$ would have resulted in slight underestimation.  The edge-on disk sample was corrected for galaxy contamination by removing edge-on disks which have luminosity within 0.2 $log(L)$ of a source from the galaxy sample from Section \ref{sec:gal_cont} scaled to the size of the Cep OB3b $A_{V}$ $<$ 3 region.

To determine the number of Class II sources $(N_{CII})$, we used the criteria from \cite{2009ApJS..184...18G}.   We counted all Class II sources with de-reddened $m_{24}$ brighter than the cutoff.  The relatively large size of the Taurus map (44~$deg^{2}$) along with the proximity of the region resulted in more contamination to the Class II sample.  To minimize the contamination, we modified the criteria for this cloud as well as the Lupus and Chamaeleon clouds which were combined with the Taurus region in our analysis. The AGN identification criteria for Taurus, Lupus, and Chamaeleon were taken from the criteria in the Appendix of \cite{2008ApJ...674..336G} and adjusted by lowering the $m_{4.5}$ threshold magnitude by 2.5 $mag$.

For each of the candidate protostars in the $A_{V}$ $<$ 3 region of Cep OB3b (i.e. our sample of likely edge-on disks), we estimated a faux bolometric luminosity using Equation \ref{eqn:relationship}.  Then, for each expected edge on disk, $N_{edge}$, we randomly selected one of the luminosities from this sample of likely edge-on disks.  We repeated this $N_{edge}$ times until we have constructed the luminosity function of edge-on disks.

\subsection{Reddened Disk Sources}
\label{sec:red_disk}
Highly reddened Class II YSOs can have colors and 24~$\mu$m magnitudes which satisfy the classification criteria for protostars. To find the amount of contamination due to these sources and estimate the luminosities which would be derived for these contaminants, we used a Monte-Carlo simulation.  This simulation randomly applied a realistic distribution of extinctions to the observed colors and magnitudes of a fiducial sample of Class II objects and then selects the objects which have artificially reddened photometry that fits the protostar selection criteria.  Since the reddening applied to each source is determined from the probability distribution of extinctions, we repeat this simulation 1000 times to sample the range of likely contamination.

The fiducial sample was constructed from Class II objects identified in low $A_{V}$ regions.  We first identified Class II objects in the low extinction ($A_{V}$ $<$ 3) region of our extinction map. The selected Class II sources were then de-reddened using the reddening law from \cite{2007ApJ...663.1069F}.  Sources with de-reddened $m_{24}$ brighter than the cutoff magnitude were the selected for the fiducial sample.  We used these objects as a fiducial sample of Class II objects, with the accompanying assumption that the colors and magnitudes of this sample is representative for all Class II objects in the cloud.

We then identified the full sample of Class II objects and protostar candidates in the $A_{V}$ $>$ 3 region.  We included protostar candidates since a fraction of the protostar candidates may be the reddened Class II sources.  Furthermore, protostar candidates are in more highly reddened locations and failure to include them would bias our distribution of $ A_{V}$ to lower values.  We sorted the Class II objects and protostar candidates (hereafter: YSOs) into regions of higher and lower stellar density by using the distance to the 4$^{th}$ nearest YSO neighbor: sources with 4$^{th}$ nearest neighbor distances less than the median value of the protostar candidate sample in a particular cloud were considered to exist in regions of higher stellar density (``high stellar density'') and those with distances greater than the median value were considered to exist in regions of lower stellar density (``low stellar density'').  We then executed the following analysis for the high and low stellar density sources independently.  This was done because high stellar density sources may be in regions of systematically higher gas column density (and reddening) compared to sources in regions of lower stellar density.

We extracted the $A_{V}$ values coincident to each YSO using the $A_{V}$ maps; this gave us two $A_{V}$ distributions, one for the high stellar density sources, and one for the low stellar density sources.  These were the maximum $A_{V}$ values, $A_{V}(max)$ at the position of a particular YSO.   Since the YSOs are embedded in the cloud, the $A_{V}$ value to that YSO is between 0 and $A_{V}(max)$.   For each YSO with a de-reddened $m_{24}$ greater than the cutoff for that cloud, we randomly selected an $A_{V}(max)$ value and set the $A_{V}$ to a value drawn from a uniform distribution of extinctions between 0 and $A_{V}(max)$.  We then randomly selected a Class II object from our fiducial sample and applied the $A_{V}$ using the reddening law from \cite{2007ApJ...663.1069F}.  To estimate the fraction of (high or low stellar density) reddened disks masquerading as protostar candidates, we then tested to see how many of the reddened Class IIs were identified as protostar candidates using our criteria. This process was repeated for 1000 iterations for the high and low stellar density sources independently.   In each of the 1000 iterations, we estimated the faux bolometric luminosities of the contaminants using the same technique used for the protostar candidates using Equations \ref{eqn:relationship} and \ref{eqn:flat}.  For each cloud, we were left with two distributions of contaminant luminosities; one for the sources in regions of high stellar density and one for the sources in regions of low stellar density.  

\begin{threeparttable}
\caption{Star Forming Region Properties}
\centering
\begin{tabular}{c c c c c c}
\hline\hline
& & & &  Protostars & Contamination \\
\hline 
Cloud & Dist. (pc) & $m_{24}$ cut & Class IIs\tnote{a} & Flat/Rising\tnote{b} & Red/Edge-on/Gal. \\
\hline 
Ophiuchus & 125\tnote{1} & 5.5 & 122 & 14(11.4)/15(8.7) &  3.3/5.6/0.0 \\
Tau/Lup/Cha & 140\tnote{2} & 5.0 & 106 & 9(6.0)/25(21.6) &  2.1/4.3/0.0 \\
Perseus & 230\tnote{3} & 6.0 & 192 & 13(10.5)/44(34.0) &  3.3/9.1/0.0 \\
Serpens & 415\tnote{4} & 7.0 & 112 & 11(7.1)/29(22.8) &  3.5/5.2/1.4  \\
Orion & 420\tnote{5} & 7.0 & 1761 & 89(68.8)/217(160.1) &  11.9/60.1/5.3  \\
Cep OB3 & 700\tnote{6} & 8.0 & 505 & 43(27.7)/105(73.9) & 5.3/32.5/8.6 \\
Mon R2 & 830\tnote{7} & 8.0 & 347 & 31(28.1)/82(67.5) &  3.1/11.5/2.8 \\
\hline
\end{tabular}
\begin{tablenotes}
\item[1]{From \cite{2009ApJS..181..321E}.}
\item[2]{Taurus distance from \cite{1994AJ....108.1872K}, Lupus I, II, $\&$ IV  and Chamaeleon II are assumed at this distance.}
\item[3]{From \cite{1990Ap&SS.166..315C}.}
\item[4]{From \cite{2010ApJ...718..610D}.}
\item[5]{From \cite{2007A&A...474..515M}.}
\item[6]{From \cite{2005A&A...438.1163K}.}
\item[7]{From \cite{1968AJ.....73..233R}.}
\item[a]{Class II sources in regions with $A_{V}$ $>$ 3 and de-reddened $m_{24}$ $<$ cutoff.}
\item[b]{Numbers in parenthesis are after removal of contamination.}
\end{tablenotes}
\label{table:sfr_prop}
\end{threeparttable}

\subsection{Contamination Removal}

Once we estimated the luminosities for the contaminating sources, we sought to remove protostar candidates with similar luminosities from our sample.  Instead of binning the data, and subtracting the luminosity function of the contaminants from that of the candidates (see Figure \ref{fig:fig8}), we eliminated individual protostar candidates from the sample using the following method.  This method allows us to create contamination subtracted cumulative distributions of the luminosities without binning.

First, we divided the sources in each cloud into high and low stellar density protostar candidates using the method described in Section \ref{sec:red_disk}. We note that this division between high and low stellar density YSOs made a difference for the reddened disk contamination, since the high stellar density regions have systematically higher extinctions, and for the galaxy contamination because likely background galaxies are assumed to be found among the lower stellar density sources, which cover a larger region of the sky.  

For each high and low stellar density sample, we generated 1000 trials of estimated contaminant luminosities for each type of contaminant: galaxies, edge-on disks and reddened Class II sources.  In each of the 1000 trials, the protostar candidate with the closest luminosity to each contaminant luminosity, such that the difference was $\delta log(L) <$ 0.2, was flagged as contamination and removed from the (high stellar density or low stellar density) protostar candidate sample.  If there was no protostar candidate which satisfied these criteria, no source was removed.  The final product is 1000 realizations of the contamination subtracted protostar sample.  

These realizations will be used to generate luminosity functions and do statistical analyses in Section \ref{sec:LFs}.  In each realization, different candidate protostars are removed as contamination.  Thus, each of the 1000 realizations will be used in the analyses that follow to take into account the uncertainty in the contamination removal.

\section{A Comparative Study of Protostellar Luminosity Functions}

After carefully selecting protostar candidates, estimating their luminosities, and removing contamination estimations, we construct luminosity functions for nine clouds, again combining the Taurus, Lupus, and Chamaeleon clouds in our sample.  These clouds cover a range of total gas masses and include both crowded clusters and regions of relatively isolated star formation  \citep{2011ApJ...739...84G,2009ApJS..181..321E}.  This provides a unique opportunity to compare populations of protostar candidates in diverse clouds and environments.  We refer to the luminosity functions of the samples of contamination subtracted protostar candidates as ``protostellar luminosity functions"; these represent our best estimation of the luminosity function of true protostars once the contamination is taken into account.  Because we have taken into account contamination, we refer to the objects in these luminosity functions as ``protostars'' instead of ``protostar candidates''.  In this section, we compare the luminosity function between clouds and within a given cloud, to examine the dependence of the luminosity function on the properties of the parental cloud and the local environment with a cloud.

\subsection{Cloud Luminosity Functions}
\label{sec:LFs}

For each of the clouds we created contamination-subtracted luminosity functions, which were generated from the 1000 realizations of the protostar candidate sample.  After each of the 1000 iterations, we determined the average number of remaining protostars per bin to generate the luminosity functions for the high and low stellar density protostars, these in turn were combined to generate one protostellar luminosity function per cloud.  We found a luminosity cutoff, $L_{cut}$, which is the $m_{24}$ cutoff magnitude translated to luminosity.  We calculated $L_{cut}$ by assuming an SED slope steeper (redder) than 90$\%$ of all sources in the region and a mid-IR luminosity equal to the 24 $\mu$m luminosity at the cutoff magnitude and the relationship in Equation \ref{eqn:relationship}.  We note that the luminosity cutoff is found from the $m_{24}$ cutoff we used to reduce the number of galaxies (Figure \ref{fig:fig3}) and is above the sensitivity cutoff.  Thus, we expect our samples to be complete down to the cutoff, except in regions with very bright nebulosity.  A more detailed description of incompleteness is given in Appendix \ref{sec:completeness} and in Section \ref{sec:intra_cloud}.

The protostellar luminosity functions and $L_{cut}$ are shown in Figure \ref{fig:fig9}.  We also show the combined luminosity functions for the clouds with high mass star formation (Orion, Cep OB3 and Mon~R2, hereafter the high mass SF clouds) and for the clouds forming low to intermediate mass stars (Serpens, Perseus, Ophiuchus, Taurus, Lupus I, III, and IV, and Chameleon II; hereafter the low mass SF clouds).   Properties of the luminosity functions, including the number of sources comprising each luminosity function, the peak, mean, median, 1~$\sigma$ values, and $L_{cut}$, are listed for each cloud as well as the combined low mass SF clouds and high mass SF clouds in Table \ref{table:LF_prop}.
 
The protostellar luminosity functions of the high mass SF clouds have peaks near 1~L$_{\odot}$ and tails extending toward higher luminosities upward of 100~L$_{\odot}$.  A combined luminosity function for these regions shows a similar peak and tail.  The median luminosity in each of the high mass SF clouds is $\sim$ 1L$_{\odot}$.  Cep OB3 has the highest mean luminosity of the high mass SF clouds, at 12.06~L$_{\odot}$.  In Orion we expect that we are missing some of the most luminous sources in the saturated regions of the Orion nebula and NGC 2024.  In all three massive clouds the most massive objects are missed due to spatially extended regions of saturation.

The low mass SF cloud luminosity functions do not show a consistent trend.  Perseus does not show a peak in the luminosity function above $L_{cut}$, but instead rises toward lower luminosities down to $L_{cut}$.  Ophiuchus shows a marginal peak in its luminosity function, but with small number statistics this peak is not significant.  The luminosity function of Tau/Lup/Cha peaks near 1~L$_{\odot}$, similar to the high mass SF clouds.  The Serpens luminosity function peaks at the highest luminosity, near 2.60~L$_{\odot}$.  The median protostar luminosity for most of the low mass SF clouds is below 1~L$_{\odot}$ except in Serpens, which has a median luminosity of 3.07~L$_{\odot}$.  The mean protostar luminosity ranges from 0.45 L$_{\odot}$ in Ophiuchus to 5.03~L$_{\odot}$ in Serpens.  The low mass SF clouds do not contain protostars at luminosities at or above 1000~L$_{\odot}$, and do not exhibit a distinct tail near 100~L$_{\odot}$ as in the high mass SF luminosity functions.  The combined luminosity function does not show a peak, but rises toward lower luminosities down to $L_{cut}$, a feature akin to the Perseus luminosity function.

\begin{center}
\begin{threeparttable}
\caption{Properties of Protostellar Luminosity Functions in $log(L/L_{\odot}$).}
\begin{tabular}{c c c c c c c}
\hline\hline
Region & Number\tnote{$\dagger$} & Peak\tnote{*} & Median\tnote{*} & Mean\tnote{*} & 1$\sigma$\tnote{*} & $L_{cut}$ \\
\hline 
Ophiuchus & 20 & -1.08\tnote{a} (-0.58\tnote{a} ) & -0.88(-0.58) & -0.76(-0.50) & 0.57(0.66) & -2.06 \\
Tau/Lup/Cha & 28 & -0.08\tnote{a} (-0.08\tnote{a} ) & -0.16(-0.10) & -0.13(-0.06) & 0.70(0.71) & -1.66 \\
Perseus & 43 & -\tnote{b} (-\tnote{b} ) & -0.46(-0.42) & -0.32(-0.25) & 0.66(0.66) & -1.30 \\
Serpens & 31 & 0.41\tnote{a} (-\tnote{b} ) & 0.49(0.25) & 0.17(0.36) & 0.75(0.85) & -1.36 \\
Orion & 229 & -0.08\tnote{a} (-0.08\tnote{a} ) & 0.06(0.22) & 0.23(0.36) & 0.72(0.74) & -1.43 \\ 
Cep OB3 & 100 & -0.08\tnote{a} (0.42\tnote{a} ) & 0.01(0.14) & 0.04(0.16) & 0.88(0.88) & -1.49 \\
Mon R2 & 96 & -0.08\tnote{a} (-0.08\tnote{a} ) & 0.02(0.09) & 0.09(0.20) & 0.68(0.69) & -1.34 \\
Low-mass & 122 & -\tnote{b} (-\tnote{b} ) & -0.20(-0.13) & -0.18(-0.09) & 0.72(0.78) & -1.30 \\
High-mass  & 425 & -0.08\tnote{a} (-0.08\tnote{a} ) & 0.04(0.18) & 0.15(0.28) & 0.76(0.77) & -1.34 \\
\hline
\end{tabular}
\label{table:LF_prop}
\begin{tablenotes}
\item[$\dagger$]{Average number of sources in the 1000 realizations of the luminosity function after contamination subtraction.}
\item[*]{De-reddened luminosity function values are given in parenthesis.}
\item[a]{Peak at center of peak bin.}
\item[b]{No significant peak above $L_{cut}$.}
\end{tablenotes}
\end{threeparttable}
\end{center}

To establish whether the observed differences in the luminosity functions are statistically significant, we perform a K-S test on each of the 1000 realizations of the contamination-removed luminosity functions.  We compare realization 1 of cloud 1 with realization 1 of cloud 2, realization 2 of cloud 1 with realization 2 of cloud 2, and so on (in each realization we have combined the high and low stellar density protostars into a single luminosity function).  All 1000 realizations are used to take into account the uncertainty in the contamination removal.  Table \ref{table:LF_KS} gives the median K-S probabilities for each combination.  We choose a threshold probability of 0.0027, equivalent to significance at the 3~$\sigma$ level in Gaussian statistics, to determine whether we can rule out the possibility of two realizations coming from the same parent distribution.   We find that among the high mass SF clouds, we cannot rule out that they are drawn from the same parent distribution: the median probability in the comparison of Orion and Cep OB3 is 0.0095, of Orion and Mon R2 is 0.0653, and of Cep OB3 and Mon R2 is 0.1168.  Comparison with the lower mass SF regions shows that the Serpens luminosity function is not likely from the same distribution as the Ophiuchus luminosity function.  The Ophiuchus luminosity function is not likely from the same distribution as the Serpens luminosity function, or the luminosity functions of any of the high mass SF clouds.  The Perseus luminosity function is not likely from the same parent distribution as any of the luminosity functions of the high mass SF clouds.  We also note that the comparison of the Ophiuchus and Perseus luminosity functions shows a probability close to our threshold.  The Tau/Lup/Cha luminosity function cannot be distinguished from any of the distributions; this may result from the small number of protostars in these regions.  We conclude that there are significant differences between the luminosity functions of the observed clouds, with some clouds showing relatively similar luminosity functions and others showing distinctly different luminosity functions.  Of particular interest is the differences between the high mass SF clouds and the low mass SF clouds.  The luminosity functions of high mass SF clouds peak at and extend to higher luminosities than the low mass SF clouds, which do not show a common, distinct peak in their luminosity functions.  In Figure \ref{fig:fig10} we show the resulting distribution of K-S probabilities for the combined high mass SF cloud luminosity function compared with the luminosity function from the combined low mass SF clouds.  The median probability is $log(prob)$ = -4.61. The observed differences in the protostellar luminosity functions suggests a real difference between the properties of the protostars in these two cloud environments.

\begin{center}
\begin{threeparttable}
\caption{K-S Comparison of Luminosity Functions}
\begin{tabular}{c c c c c c c c}
\hline
& & Median & probability & & & & \\
\hline\hline
 & Ophiuchus & Tau/Lup/Cha & Perseus & Serpens & Orion & Cep OB3 & Mon R2 \\
\hline 
Ophiuchus & - & 0.0043 & 0.0301 & -4.09\tnote{a} & -7.43\tnote{a} & -3.84\tnote{a} &  -6.37\tnote{a} \\
Tau/Lup/Cha & 0.0043 & - & 0.2561 & 0.0446 & 0.0413 & 0.3988 & 0.3872 \\
Perseus & 0.0301 & 0.2561 & - & 0.0032 & -4.93\tnote{a} & 0.0327 & -3.57\tnote{a} \\
Serpens & -4.09\tnote{a} & 0.0446 & 0.0032 & - & 0.3657 & 0.1828 & 0.0865 \\
Orion & -7.43\tnote{a} & 0.0413 & -4.93\tnote{a} & 0.3657 & - & 0.0095 & 0.0653 \\ 
Cep OB3 & -3.84\tnote{a} & 0.3988 & 0.0327 & 0.1828 & 0.0095 & - & 0.1168 \\
Mon R2 & -6.37\tnote{a} & 0.3872 & -3.57\tnote{a} & 0.0865 & 0.0653 & 0.1168 & - \\
\hline
\end{tabular}
\label{table:LF_KS}
\begin{tablenotes}
\item[a]{Probability given as $log(prob)$}
\end{tablenotes}
\end{threeparttable}
\end{center}

\subsection{Effect of Reddening}
\label{sec:reddening}
Protostars are often found in extended regions of high extinction that can further redden the protostars already reddened by their infalling envelopes.  \citet{2005PhDT........10G} devised a scheme for de-reddening pre-main sequence stars that extended the approach of \cite{1997AJ....114..288M} to the {\it Spitzer} wavelength bands.  This scheme takes into account sources with infrared excesses due to disks by de-reddening the stars onto the CTTS locus established in the combined 2MASS J, H, K$_{s}$ and IRAC 3.6, and 4.5 $\mu$m color space.  Although any object detected in two of the five bands can be de-reddened, this procedure may not be appropriate for protostars for two reasons. First, we cannot distinguish between foreground extinction and the extinction by the envelope.  Second, much of the light from the protostars at 1 - 4~$\mu$m is scattered by dust in the envelope, thus making the objects appear more blue.  Both of these effects will result in an overcorrection for reddening and an artificially high luminosity.

Since we cannot reliably de-reddened protostars using their observed colors, we instead calculate the reddening from the Class II objects in the vicinity of the protostars.  This approach, which was also adopted by the {\it c2d} team, assumes that the Class II objects are reddened by the same foreground reddening as the protostars, although Class II sources may not be as deeply embedded as the protostars.  This foreground reddening includes both that from the extended molecular gas exterior to the protostellar envelope, and that from the ISM between the observer and the molecular cloud.  Although the reddening may vary on smaller scales than accounted for in this technique, we can still assess the magnitude of the effect reddening has on the luminosity function.  The Class II objects can be de-reddened on the basis of the J, H, K$_{s}$, 3.6 and 4.5~$\mu$m using the technique described by \cite{2005ApJ...632..397G,2009ApJS..184...18G}.  To take into account variations in the foreground extinction due to structure in the parental molecular cloud, we adopt the median extinction of the five nearest Class II objects to each protostar.  We then use the reddening law of \cite{2007ApJ...663.1069F} to de-redden the photometry of the protostar candidates.  After de-reddening, we re-calculate the SED slope and $L_{MIR}$, and find bolometric luminosities.  The distribution of the ratio of de-reddened bolometric luminosity to uncorrected bolometric luminosity for all clouds is shown in Figure \ref{fig:fig11}.  We create luminosity functions using the de-reddened luminosities.  The peak, median, mean, and 1$\sigma$ are listed in parenthesis in Table \ref{table:LF_prop}.  K-S tests give the likelihood that the uncorrected and de-reddened luminosity functions are from the same parent distribution with probabilities that range from 0.08 in Orion to 0.72 in Mon R2.  This shows that although in certain regions it can be important, in most cases the reddening doesn't seem to have a large impact. Since the reddening corrections for the protostars are uncertain, we use both the uncorrected and de-reddened luminosity functions in the following analysis.  We determine a de-reddened $L_{cut}$ by adding the median increased luminosity (from uncorrected to de-reddened) to the uncorrected $L_{cut}$ for each cloud.  These de-reddened $L_{cut}$ values are (in $log$(L/L$_{\odot}$)): -1.82, -1.59, -1.21, -1.24, -1.33, -1.37, and -1.26 for Ophiuchus, Tau/Lup/Cha, Perseus, Serpens, Orion, Cep OB3, and Mon R2, respectively.

\subsection{Flat vs Rising SED protostars}
\label{sec:slope}

We identified protostar candidates as sources with a flat or rising spectrum in the mid-IR and determine the bolometric luminosity of these sources using Equation \ref{eqn:flat}.  There has been some question of the relationship between flat spectrum sources and rising spectrum sources.  Some flat spectrum sources may be rising spectrum sources observed from a face-on orientation through which emission from the warm inner layers can escape through the outflow cavity \citep{1994ApJ...434..330C, 2003ApJ...591.1049W}.  Alternatively, protostars resulting from the collapse of a flattened sheet-like cloud can also give a flat SED \citep{2006ApJ...648..484H}.  Finally, sources with tenuous envelopes can give a flat spectrum, in part because of the backwarming of the envelope \citep{1993ApJ...412..761N}; such sources may be protostars at the later stages of envelope dissipation.   \cite{2007ApJ...669..493W} also find that some flat spectrum sources may be reddened disks, although we have accounted for those in our sample.

We define flat spectrum sources as protostars with $\alpha$ between -0.3 and 0.3.  The fraction of {\it Spitzer} protostars which are flat spectrum ranges from 48\% in Ophiuchus to 23\% in Perseus.  We compare the luminosity functions of the rising and flat spectrum sources in each cloud by using K-S tests to determine the probability that the luminosity functions are drawn from the same parent distribution.  In each case, we only compare the distribution above the luminosity cutoff set for each region.  This analysis yields the following probabilities for each region: Orion (0.95), Cep OB3 (0.01), Mon R2 (0.33), Serpens (0.88), Perseus (0.81), Ophiuchus (0.34), and Tau/Lup/Cha (0.02). Thus, we find the distributions for the flat and rising stars statistically indistinguishable.  Given the similarity of the luminosity functions, and given that the sample of flat spectrum sources may contain many rising sources observed at a face-on orientation,  we only analyze the combined luminosity function for each set of sources.

\subsection{Comparison of {\it c2d} Cloud Luminosity Functions}
\label{sec:c2d_compare}

We compare the luminosity function of the 120 {\it c2d} protostars in this work with the observed protostellar luminosity function of 112 {\it c2d} protostars presented in \cite{2010ApJ...710..470D}.  For this comparison, our Serpens, Perseus, Ophiuchus, Lupus, and Chamaeleon protostar luminosity functions were combined into a single {\it c2d} cloud protostellar luminosity function.  The combined luminosity function for the {\it c2d} clouds from this work does not show a significant peak but increases toward lower luminosities and does not extend above 100L$_{\odot}$.  Similar to our luminosity function, the observed protostellar luminosity function of the {\it c2d} protostars from \cite{2010ApJ...710..470D} also does not extend above 100 L$_{\odot}$.  However, in contrast to our luminosity function, it shows a distinct peak above 1L$_{\odot}$.  Thus, we find more low luminosity protostars in the {\it c2d} clouds than \cite{2010ApJ...710..470D}.

\cite{2011ApJ...736...53O} give dimensionless quantities for the observed luminosity function from \cite{2010ApJ...710..470D} in their Table 3, particularly the ratio of the median bolometric luminosity to the mean bolometric luminosity,  $L_{bol}^{med}$/$L_{bol}^{mean}$ = 0.3, and the standard deviation of $log(L_{bol})$, $\sigma$(logL) = 0.7.   For our combined {\it c2d} luminosity function, $L_{bol}^{med}$/$L_{bol}^{mean}$ = 0.22 and $\sigma$(logL) = 0.72.  These values are the same as those derived for the \cite{2010ApJ...710..470D} luminosity function.  De-reddening the photometry using the method described in Section \ref{sec:reddening} brings $\sigma$(logL) to 0.77, which is consistent with the value for the \cite{2010ApJ...710..470D} luminosity function, and $L_{bol}^{med}$/$L_{bol}^{mean}$ to 0.15, which is less than the value for the \cite{2010ApJ...710..470D} luminosity function.  Thus, we find that on the basis of the quantitative statistics used by \cite{2011ApJ...736...53O}, our {\it c2d} luminosity function has a similar width as the luminosity function from \cite{2010ApJ...710..470D}, but when de-reddened, our luminosity function has a lower $L_{bol}^{med}$/$L_{bol}^{mean}$.  The main difference is that we find an excess of faint objects and no peak.  The reason for the difference may be that \cite{2010ApJ...710..470D} require 1.3 mm envelope detections for their protostars; this may remove very low mass protostars with envelope masses less than 0.5 M$_{\odot}$, from being included in their protostar sample \citep{2009ApJS..181..321E}.

\subsection{Comparing Protostars in Regions of High and Low Stellar Density}
\label{sec:intra_cloud}

Molecular clouds host a variety of star forming environments, including regions of high and low stellar density YSOs.  Although it has been shown that OB stars are typically found in clusters \citep{1999A&A...342..515T}, it is unclear whether this is because of the fact that massive stars are rare and thus are only likely to be found in groups of low mass stars \citep{2006MNRAS.370..488B}, or because these stars preferentially form in clusters \citep{2010A&A...524A..18B}.  Since the luminosity of a protostar is a combination of accretion luminosity and intrinsic luminosity, we cannot determine the masses of our protostar candidates; however the most luminous protostars tend to be the most massive protostars \citep{2003ApJ...585..850M}.  We now compare the luminosity functions of high and low stellar density protostars to determine if the luminous protostars are found preferentially in high stellar density environments.

To provide a measure of the YSO density around each protostar candidate, we computed nearest-neighbor distances.  The nearest-neighbor (nn) distance is the distance to the $n^{th}$ nearest Class II or protostar candidate.  We chose n = 4 after considering n = 2 and  n = 10 distances; n = 4 gives a better indication of clustering in both low and high stellar density star forming clouds while n = 2 is dominated by random fluctuations \citep{1985ApJ...298...80C} and n = 10 is not sensitive to clustering in smaller groups (hereafter, the distance to the $4^{th}$ nearest YSO will be denoted ``nn4 distance'').

We show in Figure \ref{fig:fig12} the bolometric luminosity vs.~nn4 distance for all protostar candidates.  For the three high mass SF clouds, Orion, Cep OB3, and Mon R2, a trend can be noted between the nn4 distance and luminosity of the most luminous protostar at that nn4 distance.  In these clouds, as the nn4 distance decreased the luminosity of the most luminous protostar increased.  This can also be seen in the combined plot for all three high mass SF clouds.  Thus, the most luminous sources ($L > 10~L_{\odot}$) are typically found with nn4 $<$ 0.50 pc.  

This trend is more clear for the Orion cloud, in which we have the largest sample of protostar candidates.  However, the most luminous protostar candidate in the Orion sample has an nn4 distance of 0.52 pc, greater than the median nn4 distance.  In addition, the fifth most luminous protostar candidate in Orion has an nn4 distance of 0.38 pc, greater than the typical distance for sources of comparable luminosity in the Orion cloud.  These sources, Reipurth 50 and V883~Ori, are thought to be undergoing FU~Ori eruptive events \citep{1993ApJ...412L..63S}.  Thus, these sources appear to be low mass protostars undergoing luminous outbursts in which the accretion luminosity dominates the intrinsic luminosity \citep{1996ARA&A..34..207H}.  Additionally, an outlier is seen in the third most luminous protostar candidate in Cep OB3 at an nn4 distance of 0.47 pc.  We speculate that this may be an outburst source as well.

The low mass SF clouds do not show the same trend.  Instead, we find no convincing correlation between the most luminous protostar candidates and nn4 distance.  We perform a two-dimensional K-S test comparing the distributions of nn4 distance and luminosity for protostar candidates with luminosity above the most conservative $L_{cut}$ in the high mass SF clouds with those in the low mass SF clouds.  We find a probability of 0.0023 that these distributions might be from the same parent distribution, which is below our threshold.  The low mass SF clouds are different from high mass SF clouds because they contain few sources above 10~L$_{\odot}$.  In addition, the broad range of nn4 distances found in the high mass SF clouds is not apparent in most of the low mass SF clouds.  

Is the trend that we see in Figure \ref{fig:fig12} the result of a decreasing sample size at larger nn4 (and hence fewer of the rare luminous protostars) or is it due to a real change in the luminosity function?  To test this, we separate the Orion protostars into two equal-sized samples based on nn4 distance.  The two samples are separated using the median protostar candidate nn4 distance, $D_{c}$, such that one sample have nn4 distances less than $D_{c}$, and the other sample have nn4 distances greater than $D_{c}$.  Then, we compare the luminosity functions of protostars in regions of higher stellar density (nn4 distances $<$ $D_{c}$) and luminosity functions of protostars in regions of lower stellar density (nn4 distances $>$ $D_{c}$).  In Figure \ref{fig:fig13} we show the luminosity functions of the protostars in regions of high and low stellar density in Orion.  The left panel shows the distributions using the $m_{24}$ cutoff implemented in this work; the right panel emphasizes the small effect a one magnitude difference in the $m_{24}$ cutoff would have on completeness and is discussed in further detail below.  The luminosity function of the protostars in regions of high stellar density peaks above 1 L$_{\odot}$, while the luminosity function of the protostars in regions of lower stellar density peaks at 1 L$_{\odot}$.  We see for the contamination-subtracted luminosity functions, as in Figure \ref{fig:fig12} for all protostar candidates, the majority of sources above 10 L$_{\odot}$ are in regions of higher stellar density, though again we see the most luminous protostar (the potential outbursting source) is indeed in a region of low stellar density.

To test whether the differences between the high and low stellar density populations are significant, we perform 1000 K-S tests comparing the contamination-removed luminosity functions of the high and low stellar density regions in each cloud using the same method as the cloud-to-cloud comparison.  The results are shown in Table \ref{table:KS_clust_iso} for both the uncorrected and de-reddened luminosity functions.  Listed are the median K-S probabilities for the 1000 trials that high and low stellar density protostar luminosity functions are from the same parent distribution.  For Orion, all 1000 probabilites are shown in Figure \ref{fig:ks_clust_iso}.  We again use the threshold of 0.0027 to determine if we can rule out the possibility of the high and low stellar density luminosity functions coming from the same parent distribution.  We find that for both the uncorrected and de-reddened Orion luminosity functions we would rule out the possibility that the luminosity functions in the high and low stellar density regions come from the same parent distribution. 

\begin{center}
\begin{threeparttable}
\caption{Comparison of high and low stellar density protostars.}
\begin{tabular}{c c c c} 
\hline\hline
Cloud & P($D_{c}^{uncorr})$ & P($D_{c}^{de-red})$ & $D_{c}$ (pc)  \\
\hline
Ophiuchus & 0.030 & 0.135 & 0.08  \\
Tau/Lup/Cha & 0.807 & 0.512 & 0.39 \\
Perseus & 0.453 & 0.690 & 0.15 \\
Serpens & 0.053 & 0.039 & 0.13  \\
Orion  & -3.04\tnote{*} & -4.23\tnote{*}  & 0.19  \\
Cep OB3 & 0.126 & 0.194 & 0.25  \\
Mon R2 & 0.042 & 0.037 & 0.22 \\
\hline
\end{tabular}
\begin{tablenotes}
\item[*]{Probability given as $log(prob)$}
\end{tablenotes}
\label{table:KS_clust_iso}
\end{threeparttable}
\end{center}

Could the difference observed between the high and low stellar density populations in the Orion cloud result from biases due to saturation and/or incompleteness in the 24~$\mu$m band?  For many of the saturated sources, we are able to recover photometry using the method described in Section \ref{sec:saturation}.  We are missing sources in the regions of extended saturation in the Orion nebula and NGC~2024 nebula.   It is in these high stellar density regions where the most luminous sources and densest clusters are found.  We expect that the inclusion of these regions would enhance the differences we see between high and low stellar density environments in the Orion cloud.  

However, incompleteness may still affect the result.  There is a bias toward recovering fainter sources in less nebulous regions than in more nebulous regions because of the mid-IR nebulosity that exists most prominently in the {\it MIPS} 24~$\mu$m images.   Source crowding will affect our sample less since we have used PSF photometry extraction at 24~$\mu$m.  Appendix \ref{sec:completeness} discusses the completeness on our sample.  This shows that in both the high and low stellar density regions we do not expect to detect all protostars down to our luminosity cutoff.  The bright nebulosity and sources common in high stellar density regions may alter the luminosity function in these regions by preferentially hiding the faintest protostars in those regions.  

One way of addressing the issue of incompleteness is to change the limiting $m_{24}$ used in the Orion luminosity function. Since the incompleteness comes primarily at the faintest magnitudes (Appendix \ref{sec:completeness}), changing the magnitude limit should reduce the effect of incompleteness.  In Figure \ref{fig:fig13}, we show the average of 1000 iterations of low and high stellar density luminosity functions after adopting a limiting magnitude of $m_{24}$ $<$ 6 $mag$ and $m_{24}$ $<$ 7 $mag$.  We find the median K-S probability that the high and low stellar density protostar luminosity functions are from the same parent distribution to be 0.0013 for the original $m_{24}$ $<$ 7 $mag$ cutoff.  If we use a cutoff of $m_{24}$ $<$ 6 $mag$, the median probability is 0.0015.  Thus the result that the high and low stellar density luminosity functions are significantly different does not change after raising the limiting $m_{24}$ by one magnitude.  

Another way to assess the incompleteness is to compare the distribution of $m_{24}$ for Class II sources in high and low stellar density regions in Orion.  If our sample is incomplete in regions of high or low stellar density, we expect to find differences between distributions of $m_{24}$ of Class II sources in these two regions.  We use $m_{24}$ since we cannot use our conversion to luminosity for Class II objects.  We identify Class II sources in Orion which are within $D_{c}$ of a protostar candidate, then classify the Class II source as being in a region of high stellar density if the nearest protostar candidate is in a high stellar density region or classify the Class II source as being in a region of low stellar density if the nearest protostar candidate is in a low stellar density region.  Figure \ref{fig:fig15} shows the $m_{24}$ distribution for high and low stellar density regions.  A K-S test gives a 0.61 probability that the distributions of Class II sources in regions of high and low stellar density are from the same parent distribution.  We note that the $m_{24}$ histogram for the protostars in high and low stellar density regions of Orion show the same difference as the luminosity functions: the histogram for the high stellar density regions extends to brighter magnitudes with only a probability of 0.0089 that the two distributions are drawn from the same parent population.  If we remove the two outbursting sources this probability drops to 0.0051.  Although these are above our 0.0027 threshold, they still correspond to the probabilities equivalent to 2.6 $\sigma$ detection in Gaussian statistics.  We conclude that the differences we see between $m_{24}$ distributions of protostar candidates in regions of high and low stellar density are not due to incompleteness in $m_{24}$ because we do not see differences in the $m_{24}$ distributions of the sample of Class II sources in regions of high and low stellar density.

The same trend of increasing luminosity with decreasing nn4 distance is also apparent for Mon R2.  The de-reddened Mon R2 luminosity function shows a probability that the low and high density regions come from the same parent distribution is equal to our threshold; this is further evidence that the luminosity functions of the high and low density regions differ.  The trend is less distinct for Cep OB3.  The Cep OB3 low and high density luminosity functions are statistically indistinguishable.  This may be due to the lower number of protostars at high stellar density (as evidenced by the $D_{c}$).  Nevertheless, this cloud contains several more luminous objects that appear to be widely isolated; these objects should be followed up in future studies to determine if they are undergoing outbursts.

\section{Discussion}

\subsection{Comparison with Model Luminosity Functions}

The sample of protostellar luminosity functions  presented in this paper provide the means to examine protostellar evolution in a diverse set of nearby clouds.  The luminosity of a protostar is a combination of the intrinsic luminosity of the central protostar and the luminosity generated by accretion.  Since much of the total protostellar luminosity is derived from accretion, the luminosity functions can put constraints on the rate of accretion onto the central star.  This in turn can constrain models that determine the rate of infall of material on the central disk, and the subsequent accretion of matter from the disk onto the star.  For an initial look as to how our data may constrain these models, we compare our results to recent models of protostellar luminosity functions in the literature.

\cite{2010ApJ...710..470D} compared the {\it c2d} luminosity function with models of protostars based on a singular isothermal sphere collapse.  This comparison includes five models which extend the analysis of \cite{2005ApJ...627..293Y} by incorporating the following: scattering, an axisymmetric disk, an axisymmetric envelope, outflows and mass loss, and episodic accretion.  In each case, except the episodic accretion (EA) model, the resulting model luminosity function peaks above the observed luminosities presented in \cite{2010ApJ...710..470D} and in our sample.  One explanation for the lower observed luminosities is that the infalling material builds on protostellar accretion disks, and that the material on the disk is episodically dumped onto the protostar creating a large, but brief, jump in the source luminosity \citep{1995ApJS..101..117K,2010ApJ...710..470D}.  \cite{2010ApJ...710..470D} find that models which include episodic accretion can reproduce the range of luminosities observed in their {\it c2d} luminosity function.  

Figure \ref{fig:fig16} shows our combined low mass SF cloud luminosity function and the luminosity function from the EA model.  This model has been designed for the low mass star-forming {\it c2d} regions with a mass limit of 3 M$_{\odot}$, and thus we do not compare this model with the high mass SF clouds.  The range of luminosity produced by the episodic accretion model matches well with our low mass SF cloud luminosity function.  However, the \cite{2010ApJ...710..470D} model has a peak near 5.5 L$_{\odot}$, while the low mass SF cloud luminosity function from this work has no significant peak above $L_{cut}$.  De-reddening (see Section \ref{sec:reddening}) does not produce a significant peak in the low mass SF cloud luminosity function nor improve the match between the low mass SF cloud luminosity function and the EA model.  We thus find that our luminosity function does not exhibit the peak evident in the EA model from \cite{2010ApJ...710..470D}.

\cite{2011ApJ...736...53O} compared a variety of accretion models to the observed luminosity functions of the protostars from \cite{2009ApJS..181..321E}, including isothermal sphere, turbulent core, two-component turbulent core, competitive accretion, and two-component competitive accretion models with both accelerating and non-accelerating star-formation rate, and tapered or untapered accretion.  We compare our observed luminosity functions to the Offner $\&$ McKee models in three ways.  First, we again use the ratio $L_{bol}^{med}$/$L_{bol}^{mean}$ and the standard deviation $\sigma$(logL) (see Section \ref{sec:c2d_compare}) to compare our observed luminosity functions with the corresponding values for the models from \cite{2011ApJ...736...53O}. The values for the low mass SF clouds come from Table 3 of \cite{2011ApJ...736...53O}, which use an upper mass limit of 3 M$_{\odot}$; values for the high mass SF clouds use an upper limit of 10 $M_ {\odot}$ and are listed in Table \ref{table:model_plf} (S. Offner, P. Com). Second, we present a visual comparison of the data and models in Figure \ref{fig:fig17}.  Finally, in Table \ref{table:model_KS}, we show the KS test result of the comparison of the models to the data.  We do not provide a comparison to the isothermal sphere model to our luminosity functions since it clearly does not match any of our observed luminosity functions \citep{2011ApJ...736...53O}.

\begin{center}
\begin{threeparttable}
\caption{Protostar Luminosity Function Statistics}
\begin{tabular}{l c c c c} 
\hline\hline
& High Mass SF & & Low Mass SF & \\
\hline
Model & $L_{bol}^{med}$/$L_{bol}^{mean}$ & $\sigma$(logL) & $L_{bol}^{med}$/$L_{bol}^{mean}$ & $\sigma$(logL) \\
\hline
Turbulent Core\tnote{*} & 0.16 & 0.73 & 0.42 & 0.77 \\
Two-Component Turbulent Core\tnote{*} & 0.24 & 0.60 & 0.60 & 0.55 \\
Competitive Accretion\tnote{*} & 0.11 & 0.77 & 0.21 & 0.81 \\
\hline
Observed\tnote{$\dagger$} & 0.11 & 0.76 & 0.23 & 0.72 \\
Observed, de-reddened\tnote{$\dagger$} & 0.11 & 0.77 & 0.17 & 0.76\\
\hline
\end{tabular}
\begin{tablenotes}
\item[*]{Low mass SF clouds from Table 3 of \cite{2011ApJ...736...53O}, high mass SF cloud values from P. Com with S. Offner.}
\item[$\dagger$]{Values of the contamination-subtracted combined luminosity functions shown in Figure \ref{fig:fig9}.}
\end{tablenotes}
\label{table:model_plf}
\end{threeparttable}
\end{center}

The competitive accretion (CA) model shows a good match to $L_{bol}^{med}$/$L_{bol}^{mean}$ and $\sigma$(logL) for the low and high mass SF clouds. The CA model also provides a reasonable match to the high and low mass luminosity functions in Figure \ref{fig:fig17}.  Compared to the other models, the peak of the CA model is better matched to the broad plateau of the low mass regions and it is well matched to the luminosity of the peak in the high mass regions, particularly for the de-reddened luminosity functions, although the model under-predicts the amplitude of the peak observed toward the high mass regions.  The KS test probabilities show that the observed high and low mass luminosity functions have the highest probability of being drawn from the CA model, although there is still a low probability that the high mass luminosity function, particularly the uncorrected luminosity function, is drawn from the CA model.

The turbulent core (TC) model provides a reasonable match for the $L_{bol}^{med}$/$L_{bol}^{mean}$ and $\sigma$(logL) of the high mass SF clouds, but overestimates $L_{bol}^{med}$/$L_{bol}^{mean}$ by a factor of two for the low mass regions.  This is apparent in Figure \ref{fig:fig17}, where the this model shows a strong peak which is not apparent in the low mass star forming regions.  On the other hand, it provides a good match to the high mass regions, although the peak is at too high a luminosity.  The observed high and low mass luminosity functions shows a significantly lower probability of being drawn from the TC model than from the CA model.

The two component turbulent core (2CTC) model provides a poor match to $L_{bol}^{med}$/$L_{bol}^{mean}$ and $\sigma$(logL) of the high high and low mass SF cloud sample.  The luminosity distributions of the 2CTC models peak at higher luminosities than the high and low mass regions, and the peak of this model is much sharper than apparent in the low mass SF clouds.  This is reflected in the K-S probabilities, which show a very low probability that our data are drawn from this model luminosity function.

\begin{center}
\begin{threeparttable}
\caption{K-S Comparison of Luminosity Functions to Models}
\begin{tabular}{c c c c c}
\hline\hline
Regions & CA\tnote{a} & 2CTC\tnote{a} & TC\tnote{a} & EA\tnote{a} \\
\hline
Low Mass SF & 0.2562 & -4.60\tnote{b} & 0.0234 & -16.08\tnote{b} \\
High Mass SF & -5.36\tnote{b} & -12.92\tnote{b} & -9.26\tnote{b}& - \\
\hline
Low Mass SF, de-reddened & 0.6837 & -3.59\tnote{b}  & 0.0656 & -16.35\tnote{b}\\
High Mass SF, de-reddened & 0.0075 & -7.23\tnote{b} & -4.66\tnote{b} & - \\
\hline
\end{tabular}
\label{table:model_KS}
\begin{tablenotes}
\item[a]{Determined for luminosities above 0.1~L$_{\odot}$.}
\item[b]{Probability given as $log(prob)$}
\end{tablenotes}
\end{threeparttable}
\end{center}

Another two-component model is analyzed by \cite{2011ApJ...743...98M}, particularly a model similar to the two-component turbulent core model of \cite{2003ApJ...585..850M} and the two-component thermal and non-thermal model of \cite{1992ApJ...396..631M}.  \cite{2011ApJ...743...98M} find that this model reproduces a realistic IMF between 0.1 and 10 M$_{\odot}$.  The model luminosity function shows a peak near 1 L$_{\odot}$, akin to the luminosity functions from our sample of clouds which form high mass stars. Myers (in prep) compare the resulting luminosity function with our Orion protostellar luminosity function and find good agreement.

In conclusion, the shape of the luminosity functions of our high mass and low mass SF clouds are best matched by those produced by the competitive accretion luminosity function of \cite{2011ApJ...736...53O}.  We note that there are still significant differences between this luminosity function and the observed luminosity functions.  We will further discuss the implications of this in Section \ref{sec:wrapup}.

\subsection{The Luminosity Function and Primordial Mass Segregation}

Of particular interest is whether the different luminosity functions imply a difference in the distribution of mass of the emerging stars forming in high and low stellar density regions.  This would be a form of primordial mass segregation, with potentially more massive stars forming preferentially in denser regions.  We stress again that the bolometric luminosities we estimate are the sum of the intrinsic luminosity of the protostar and the accretion luminosity resulting from gas falling onto the protostar.  In low mass protostars, the accretion luminosity probably dominates, but for higher mass objects the intrinsic luminosity becomes increasingly important \citep{2003ApJ...585..850M,2011ApJ...736...53O}.  We make no attempt to estimate the contribution of each of these luminosity sources, and thus, it is not clear whether the most luminous protostars are those with the highest intrinsic luminosity or whether they have the highest accretion rates.  If they are sources with a higher intrinsic luminosity, then these sources probably are higher in mass, particularly if they follow a stellar birthline \citep{1997ApJ...475..770H,1991ApJ...375..288P}.  In the case that the luminosity is dominated by accretion, then a higher luminosity might imply a higher protostellar mass or a higher accretion rate \citep{2005ApJ...627..293Y,2011ApJ...743...98M}.  This might also imply a higher outcome mass for the protostar; however, if the accretion rate is higher and the accretion time is shorter, then sources in clusters may exhibit higher luminosities but have the same outcome IMF.  Thus, although one interpretation of the different luminosities is that the high stellar density regions contain protostars that are more massive or that will form stars of higher mass, this is not a unique interpretation. 

If we are indeed seeing primordial mass segregation, then this would suggest that the presence of massive stars in the dense center of clusters is due to the environment found in these crowded regions, and not the statistical sampling of a constant IMF \citep{1999ApJ...515..323E,1999MNRAS.309..461B}.  The competitive accretion model predicts a correlation between the density of stars in a cluster and the mass of the most massive member \citep{2004MNRAS.349..735B}.  This correlation is due to the large mass of gas drawn in by the gravity of the entire cluster of stars and accreted onto the most massive members of the cluster. The result is a higher mass accretion rate for the protostars in dense, clustered environments, particularly the most massive stars.  

Although we observe variations in the luminosity function with stellar density, this may in effect be a dependence between the luminosity function and the gas column density.  \cite{2011ApJ...739...84G} presented observational evidence for a Schmidt-like star formation law in molecular clouds where the star formation rate per unit surface area is proportional to the gas density, squared.  In this case, the stellar density may be just tracing the overall column density of the natal gas.  Consequentially the potential dependence of the IMF on stellar density may in fact be a dependence of the IMF on gas column density.  

A dependence between stellar density and gas density is predicted by the competitive accretion model, but could also result from Jeans fragmentation in sheet-like clouds \citep{2011ApJ...739...84G}. However, although Jeans fragmentation might explain the increasing stellar density with increasing gas density, it cannot explain the increasing stellar mass with increasing gas and stellar density since the Jeans mass decreases with increasing gas density.  An increasing mass accretion rate with increasing gas column density is predicted by turbulent core models, but it is not clear whether such models could also explain the higher stellar densities in the vicinity of higher mass stars \citep{2011ApJ...740..107C}.  Thus, the competitive model is attractive in that it can explain the observed protostellar luminosity functions and their dependence on stellar density.  

\subsection{Comments on the Comparison with Models and its Implications of the IMF}
\label{sec:wrapup}

Based on the observational study presented in this paper, we can draw the following conclusions: that the peak of the luminosity function of clouds forming high mass stars is near 1 L$_{\odot}$ while for the clouds not forming high mass stars the luminosity functions show a broad plateau that peaks below 1 L$_{\odot}$, that the luminosity functions of the high and low mass SF clouds are different, and finally, that the luminosity functions drawn from low and high density regions in Orion are different.  Our comparison of with the current models available in the literature show that none of the available models provide an excellent match to the observed luminosity functions.  This is not surprising. The Offner \& McKee models do not take into account details of protostellar evolution such as the clearing of the envelope and the possibility of some degree of episodic accretion.  On the other hand, the \cite{2010ApJ...710..470D} model only considers models staring with the collapse of a singular isothermal sphere.  We expect that these results, as well as future luminosity functions derived from Herschel data, will drive the development of realistic models of protostellar models.

Within the available models, the competitive accretion model provides the best match.  It best reproduces the shape of the low mass SF cloud and high mass SF cloud luminosity functions.  We note that this model works well since it invokes an accretion rate that is dependent on both the instantaneous and final mass of the star; this is different than models of isothermal sphere collapse where the infall rate is constant and depends only on the sound speed in the gas \citep{2011ApJ...736...53O}.  The different luminosity functions in the high and low mass SF clouds can be approximately reproduced with luminosity functions of an ensemble of protostars forming a \cite{2005ASSL..327...41C} IMF truncated at upper mass limits of 3 M$_{\odot}$ and 10 M$_{\odot}$ for the high and low mass SF clouds, respectively (note that we have excluded the regions of high mass star formation from our study of high mass star forming clouds due to saturation).  Furthermore, the variations of the protostellar luminosity function with the spatial density of YSOs is also a prediction of the competitive accretion model, where the most massive stars are formed preferentially in the centers of dense clusters of stars  \citep{2004MNRAS.349..735B}.  For these two reasons, we currently favor the competitive accretion model.  The protostellar luminosity function, however, does not provide a definitive test for models of protostellar evolution;  these models must be tested on other grounds as well, such as the measured velocities of the protostars and the density of the surrounding gas \citep{2005Natur.438..332K,2007MNRAS.374.1198A,2007ApJ...669..493W} and the distribution of sources in a bolometric luminosity and temperature diagram \citep{2010ApJ...710..470D}.  Only though comprehensive observations examining in detail the many facets of protostars and their environments, can we both test models of protostars and drive the refinement of those models to incorporate all the essential elements of protostellar evolution.

\section{Summary}
 
We identify 727 protostar candidates in {\it Spitzer} surveys of nine star forming clouds within 1 kpc of the Sun.  The sample includes both nearby dark clouds forming primarily low mass stars, clouds with moderate sized clusters forming intermediate mass stars, and clouds with large clusters forming massive stars. The clouds are Ophiuchus, Chamaeleon II, Lupus I, III, and IV, Taurus, Perseus, Serpens, Orion, Cep OB3, and Mon R2.  With this sample, we have done the following analysis:

\begin{itemize}
\item{We estimate bolometric luminosities based on SED 3 - 24 $\mu$m slope ($\alpha$) and the 1 - 24 $\mu$m luminosity, $L_{MIR}$.  We do this by determining a relationship between $L_{bol}/L_{MIR}$ and log($\alpha$) using {\it c2d} sources with known bolometric luminosities, then applying the relationship to all protostars in our sample and estimating bolometric luminosities for each of our protostar candidates.}

\item{We estimate the amount of contamination from reddened Class IIs, edge-on disks, and background galaxies.  We find that up to 20$\%$ of our protostar candidate sample are likely edge-on disks, 4$\%$ are likely reddened Class II objects, and 2$\%$ are likely galaxies.  These contaminating objects are removed from our sample using a statistical approach.}

\item{The resulting protostellar luminosity functions for clouds that for high mass stars (Orion, Cep OB, and Mon R2) peak near 1~L$_{\odot}$.  The luminosity functions of each high mass SF cloud has a tail extending toward luminosities upward of 100~L$_{\odot}$.} 

\item{The protostellar luminosity function of the low mass SF clouds (those without high mass stars) do not show a common peak.  The combined Taurus/Lupus/Chamaeleon luminosity function shows a marginal peak near 1~L$_{\odot}$, the Ophiuchus luminosity function shows a marginal peak below 1~L$_{\odot}$, Serpens shows a broad peak near 2.6~L$_{\odot}$, and Perseus do not show a well-defined peak.  None of the low mass SF clouds contain protostars at luminosities above 1000~L$_{\odot}$ nor do they show a tail above 100 L$_{\odot}$.}

\item{We find significant differences between the combined low mass SF cloud luminosity function and the combined high mass SF cloud luminosity function.  The median probability that they are from the same parent distribution is $log(prob)$ = -4.66.  Thus, we do not expect the luminosity function of protostars in high mass SF clouds are from the same parent distribution as the luminosity function of protostars in low mass SF clouds.}

\item{In the Orion clouds, there is a very low probability that the protostellar luminosity functions from high and low stellar density regions are drawn from the same parent distribution; instead the luminosity function becomes increasingly biased to higher luminosities with increasing stellar density.  This may be evidence for primordial mass segregation, although there are other possible explanations.  A similar tend is seen in Mon R2, but the trend is weak or non-existent in Cep OB3 and not present in the low mass star forming clouds.}

\item{We compare our luminosity functions with models of protostellar accretion.  We do not find a good match between the model luminosity function incorporating episodic accretion of \cite{2010ApJ...710..470D} and our low mass SF cloud luminosity function.  The combined luminosity function of both the high mass and low mass SF clouds are best matched by the competitive accretion model as implemented by \cite{2011ApJ...736...53O}, although other models cannot be formally ruled out.  Competitive accretion also predicts a dependence of the accretion rate with stellar density, consistent with the variations of the luminosity function with environment we find in our high mass SF clouds.  We conclude that models like competitive accretion which predict mass accretion rates that vary with both the mass of the protostar and the density of stars and gas in the surrounding environment, are best able to describe our observations.}
\end{itemize}

\appendix
\section{24~$\mu$m Completeness}
\label{sec:completeness}
The completeness of the protostar catalog is important to assess.  Our search for protostars in the {\it Spitzer} cloud surveys is not limited by sensitivity, but by the $m_{24}$ cutoff applied to minimize contamination from extragalactic sources. Figure 2 shows that we detect faint sources 2 magnitudes or more fainter than our $m_{24}$ cutoff.  Furthermore, at the cutoff, the number of 24~$\mu$m sources is increasing with increasing magnitude.  For these reasons, we believe our sample to be complete in most of the surveyed molecular clouds; however, because of the presence of bright nebulosity, subregions of each cloud may be affected by incompleteness.  Regions with very bright nebulosity, such as the Orion Nebula or NGC~2024, are typically saturated in the {\it MIPS} 24~$\mu$m image; these saturated regions are ignored in this paper's analysis. However, regions with signal levels well below saturation can be incomplete due to the confusion with the spatially varying nebulosity. Furthermore, regions with high stellar density can be systematically less complete due to the crowding and brighter nebulosity in clusters (Megeath et al. in prep).
 
To the assess the impact on the analysis in this paper, we concentrate on the Orion molecular clouds.  Orion contains the most luminous and nebulous star forming regions of our cloud sample.  Perhaps more importantly, Orion is the one cloud in which the luminosity functions in high and low stellar density regions are statistically different.   Thus, we need to ascertain whether the  difference between regions of high and low stellar density is real or is because of spatially varying completeness. Megeath et al. (in prep.) measure the completeness using artificial star tests in the Orion data.  They find that the completeness at a given magnitude varies with position depending on the amount of bright, saturated nebulosity.  They parameterize the amount of nebulosity using the root median square deviation, RMEDSQ = $\sqrt{median(S_{IJ}^2)-median(S_{IJ}^2)}$, which is calculated in an annulus around each source.  The RMEDSQ gives a measure of the spatially varying signal from neighboring stars and structured nebulosity surrounding each source; in the {\it Spitzer} bands the variations are dominated by the bright mid-IR nebulosity.  Using the fraction of synthetic 24~$\mu$m point sources recovered as a function of RMEDSQ. (Megeath et al in prep.), we have estimated the completeness surrounding each of the protostar candidates. The adopted RMEDSQ value is the mean value for all the YSOs within the clustering length of $D_{c}$ = 0.19 pc.  We then find the fraction of sources detected with magnitudes equal to the cutoff magnitudes.  The results of this analysis is shown in Figure 18. for both high and low stellar density regions in Orion.  For the combined high and low stellar density regions, 55\% of Orion protostars are in regions where the fraction of stars recovered is $>$ 0.90; nevertheless, both the regions of high and low stellar density have regions where the fraction of recovered sources at $m_{24}$ = 7 drops to close to 0.  We can reduce the incompleteness by reducing the cutoff magnitude.  In Figure 18, we show the same analysis for $m_{24}$ = 6.  We find that using the $m_{24}$ = 6 cutoff, 73\% are found in regions where $>$ 0.90 sources are recovered.

\begin{figure}
\centering
\includegraphics[width=0.8\textwidth]{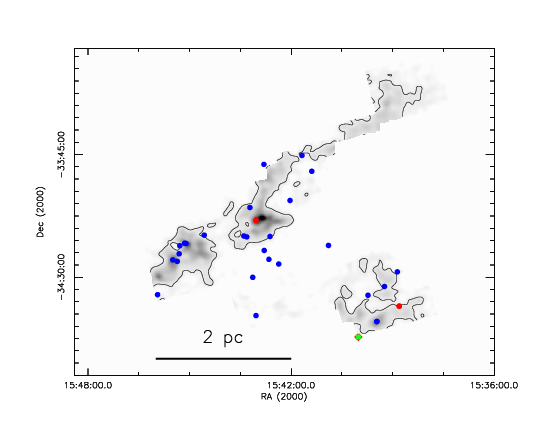}
\caption{(b). The Lupus I $A_{V}$ map.  Shown are all Class IIs (blue), rising spectrum protostar candidates  (red), and flat spectrum (green) protostar candidates with $A_{V}$ $>$ 3 (circles) and with $A_{V}$ $<$ 3 (diamonds), as well as the $A_{V}$ = 3 contours.}
\figurenum{4b}
\end{figure}

\begin{figure}
\centering
\includegraphics[width=0.8\textwidth]{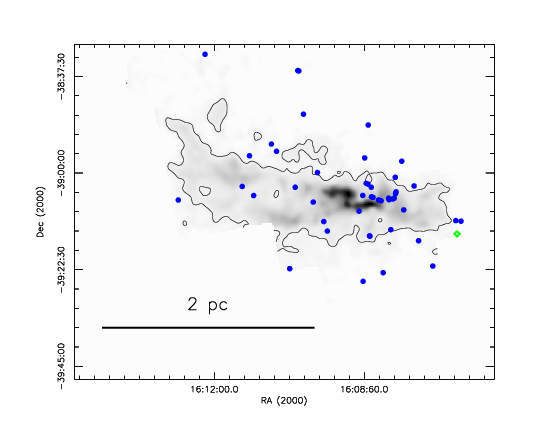}
\caption{(c). The Lupus III $A_{V}$ map.  Shown are all Class IIs (blue), rising spectrum protostar candidates (red), and flat spectrum (green) protostar candidates with $A_{V}$ $>$ 3 (circles) and with $A_{V}$ $<$ 3 (diamonds), as well as the $A_{V}$ = 3 contours.}
\figurenum{4c}
\end{figure}

\begin{figure}
\centering
\includegraphics[width=0.8\textwidth]{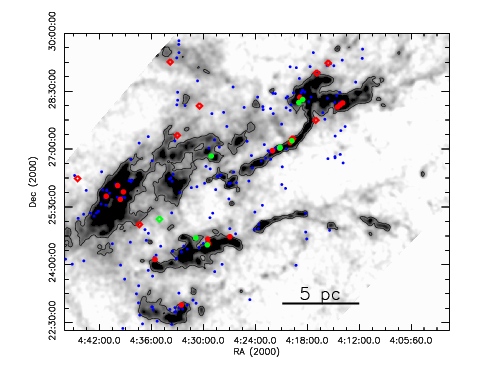}
\caption{(d). The Taurus $A_{V}$ map.  Shown are all Class IIs (blue), rising spectrum protostar candidates (red), and flat spectrum (green) protostar candidates with $A_{V}$ $>$ 3 (circles) and with $A_{V}$ $<$ 3 (diamonds), as well as the $A_{V}$ = 3 contours.}
\figurenum{4d}
\end{figure}

\begin{figure}
\centering
\includegraphics[width=0.8\textwidth]{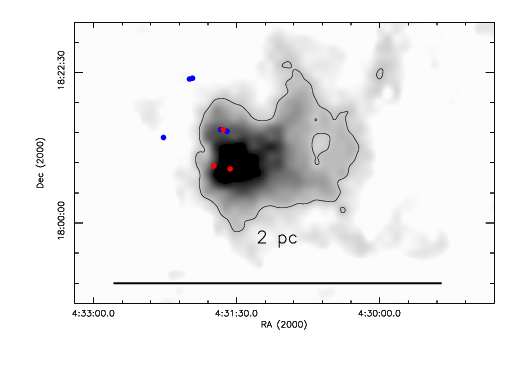}
\caption{(e.) The L1551 $A_{V}$ map.  Shown are all Class IIs (blue), rising spectrum protostar candidates  (red), and flat spectrum (green) protostar candidates with $A_{V}$ $>$ 3 (circles) and with $A_{V}$ $<$ 3 (diamonds), as well as the $A_{V}$ = 3 contours.}
\figurenum{4f}
\end{figure}

\begin{figure}
\centering
\includegraphics[width=0.8\textwidth]{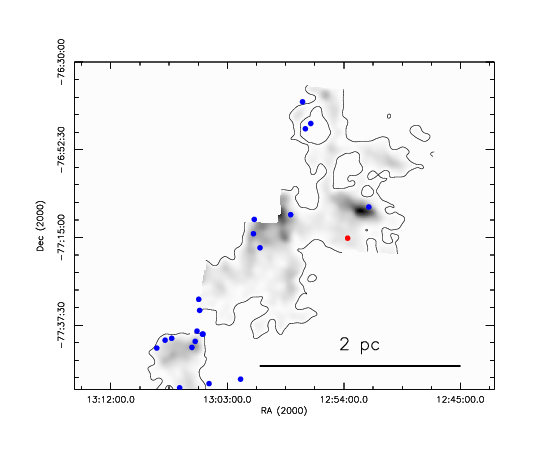}
\caption{(f). The Chamaeleon II $A_{V}$ map.  Shown are all Class IIs (blue), rising spectrum protostar candidates (red), and flat spectrum (green) protostar candidates with $A_{V}$ $>$ 3 (circles) and with $A_{V}$ $<$ 3 (diamonds), as well as the $A_{V}$ = 3 contours.}
\figurenum{4e}
\end{figure}

\begin{figure}
\centering
\includegraphics[width=0.8\textwidth]{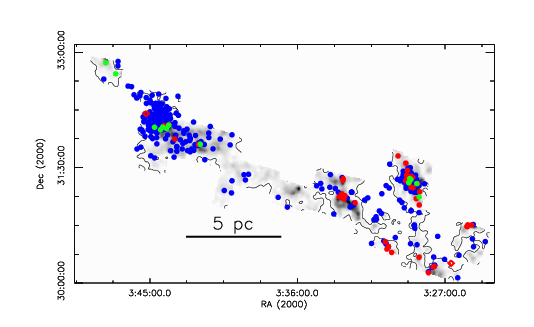}
\caption{(g). The Perseus $A_{V}$ map.  Shown are all Class IIs (blue), rising spectrum protostar candidates (red), and flat spectrum (green) protostar candidates with $A_{V}$ $>$ 3 (circles) and with $A_{V}$ $<$ 3 (diamonds), as well as the $A_{V}$ = 3 contours.}
\figurenum{4g}
\end{figure}

\begin{figure}
\centering
\includegraphics[width=0.6\textwidth]{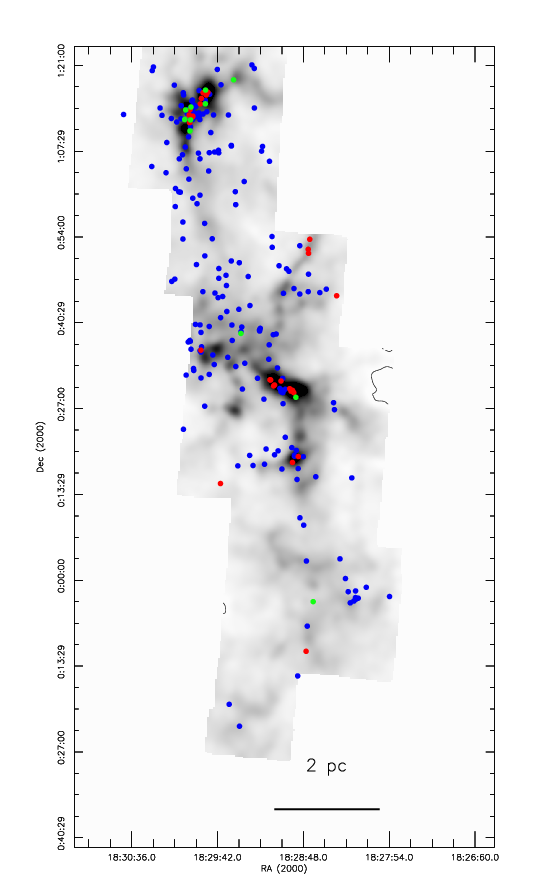}
\caption{(h). The Serpens $A_{V}$ map.  Shown are all Class IIs (blue), rising spectrum protostar candidates (red), and flat spectrum (green) protostar candidates with $A_{V}$ $>$ 3 (circles) and with $A_{V}$ $<$ 3 (diamonds), as well as the $A_{V}$ = 3 contours.  Most of the region is above $A_{V}$ = 3.}
\figurenum{4h}
\end{figure}

\begin{figure}
\centering
\includegraphics[width=0.6\textwidth]{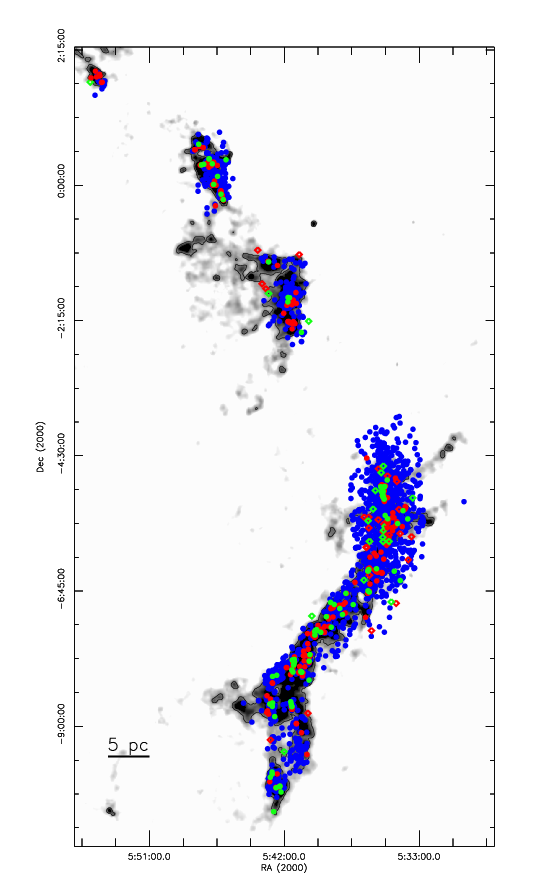}
\caption{(i). The Orion $A_{V}$ map showing both the Orion A (Dec $<$ -3.5$^{o}$) and B (Dec $>$ -3.5$^{o}$) clouds.  Our sample of Orion cloud protostars is the combined sample from both of these clouds.  Shown are all Class IIs (blue), rising spectrum protostar candidates (red), and flat spectrum (green) protostar candidates with $A_{V}$ $>$ 3 (circles) and with $A_{V}$ $<$ 3 (diamonds), as well as the $A_{V}$ = 3 contours.  The lack of YSOs near the ONC is due to saturation at 24~$\mu$m in this region.}
\figurenum{4i}
\end{figure}

\begin{figure}
\centering
\includegraphics[width=0.8\textwidth]{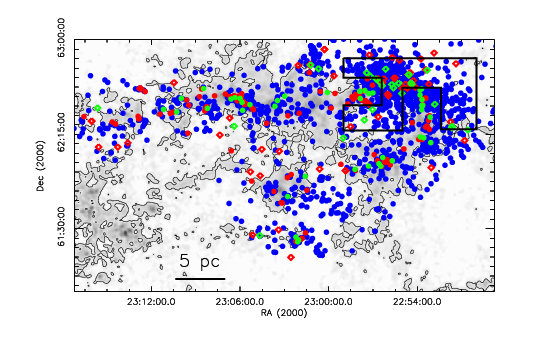}
\caption{(j). The Cep OB3 $A_{V}$ map. Shown are all Class IIs (blue), rising spectrum protostar candidates (red), and flat spectrum (green) protostar candidates with $A_{V}$ $>$ 3 (circles) and with $A_{V}$ $<$ 3 (diamonds), as well as the $A_{V}$ = 3 contours.  The $A_{V}$ $<$ 3 region used to identify likely edge-on disk contamination in Cep OB3b is also shown.}
\figurenum{4j}
\end{figure}

\begin{figure}
\centering
\includegraphics[width=0.8\textwidth]{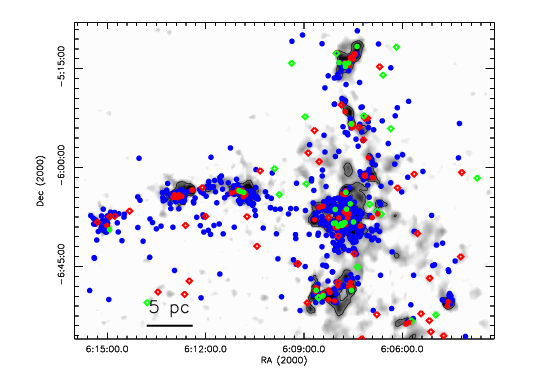}
\caption{(k). The Mon R2 $A_{V}$ map.  Shown are all Class IIs (blue), rising spectrum protostar candidates (red), and flat spectrum (green) protostar candidates with $A_{V}$ $>$ 3 (circles) and with $A_{V}$ $<$ 3 (diamonds), as well as the $A_{V}$ = 3 contours.}
\end{figure}

\begin{figure}
\centering
\includegraphics[width=0.8\textwidth]{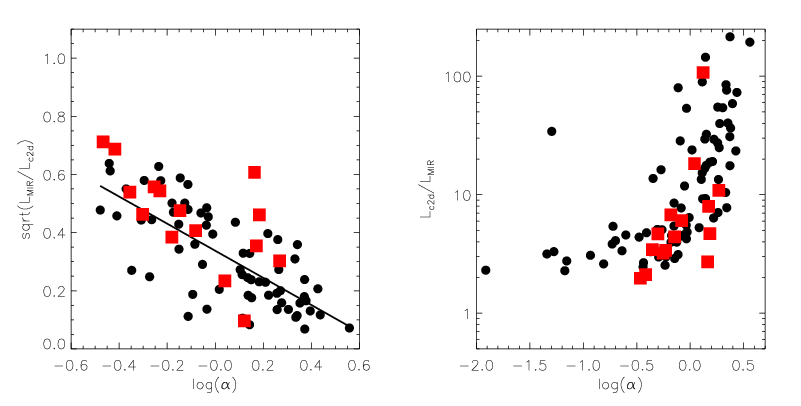}
\caption{Bolometric/Mid-IR Luminosity ratio vs. log($\alpha$) relationship.  Protostars from the {\it c2d} sample with well-constrained bolometric luminosities are plotted as black circles.  The Taurus protostars with established bolometric luminosities are plotted as red squares, but are not used to determine the fit.  Left panel: The best fit correlation (solid line) to {\it c2d} sources used to derive the relationship using rising spectrum protostars: the square-root of the fraction of mid-IR to the bolometric luminosity vs. log($\alpha$).  Right panel: The fraction of bolometric to mid-IR luminosity as a function of log($\alpha$).  The solid line shows the fit for the rising spectra protostars and adopted ratio for the flat spectrum sources at log($\alpha$) $<$ -0.5, which are included in this plot.}
\label{fig:fig5}
\end{figure}

\begin{figure}
\centering
\includegraphics[width=0.8\textwidth]{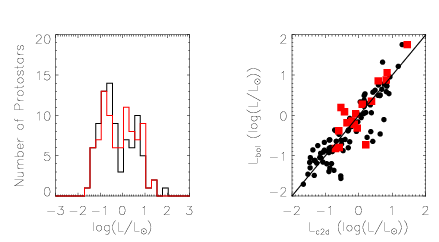}
\caption{Left panel:  Luminosity functions using estimated bolometric luminosities for {\it c2d} sources from this work (black) and actual bolometric luminosities from \cite{2009ApJS..181..321E} (red).  The {\it c2d} sources are the same used to fit the relationship in Figure \ref{fig:fig5}.  These distributions look similar, and a K-S test gives the probability that they are from the same parent distribution as 0.83.  Right panel: Shown in black are estimated luminosities from this work vs. the luminosities from \cite{2009ApJS..181..321E}.  Shown as squares (in red) are Taurus sources from \cite{2008ApJS..176..184F}.}
\label{fig:fig6}
\end{figure}

\begin{figure}
\centering
\includegraphics[width=0.8\textwidth]{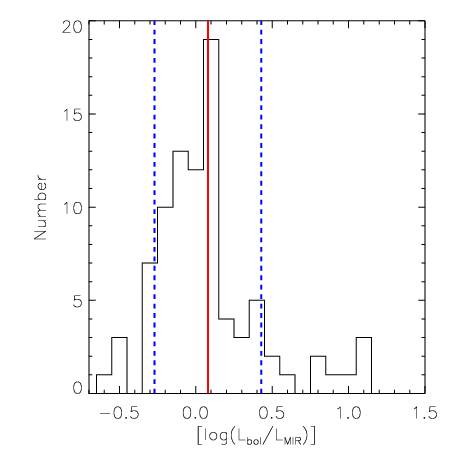}
\caption{Differences between {\it c2d} protostar bolometric luminosities estimated from our relationship and the actual $L_{bol}$ from  \cite{2009ApJS..181..321E}.  The average is shown in red with $\pm$ 1 $\sigma$ limits shown in as blue dashed lines.}
\label{fig:fig7}
\end{figure}

\begin{figure}
\centering
\includegraphics[width=0.8\textwidth]{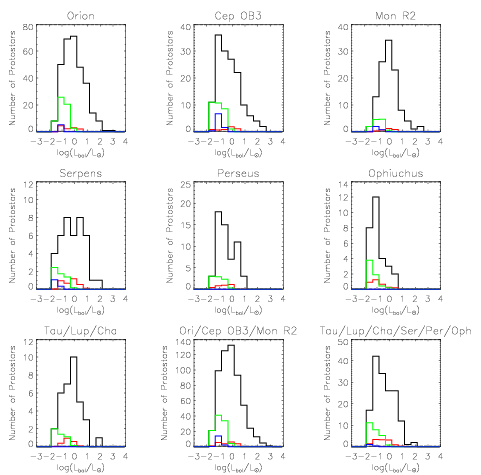}
\caption{Protostellar luminosity functions for the protostellar candidates in each of the clouds (black).  The color histograms show the contamination: reddened Class II contamination (red), edge-on Class IIs (green), and star forming galaxy (blue) contamination.  The majority of the contamination is from edge-on Class II sources, mostly falling in the lower (L $< 1 L_{\odot}$) luminosity bins.  We note that the level of edge-on Class II contamination may be overestimated.}
\label{fig:fig8}
\end{figure}

\clearpage

\begin{figure}
\centering
\includegraphics[width=0.8\textwidth]{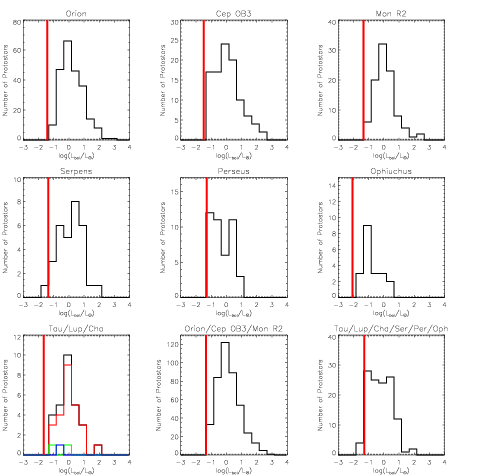}
\caption{Calculated bolometric luminosities for each region with the estimated contamination from reddened disk sources, edge-on Class IIs, and background galaxies removed.  The vertical line shows the limiting $L_{bol}$ based on the $m_{24}$ cutoff.  The panel showing combined Taurus, Lupus, and Chamaeleon luminosity function also shows the components from each cloud: Taurus is the thin red histogram, Lupus is the green, and Chamaeleon is shown as the blue histogram.}
\label{fig:fig9}
\end{figure}

\begin{figure}
\centering
\includegraphics[width=0.8\textwidth]{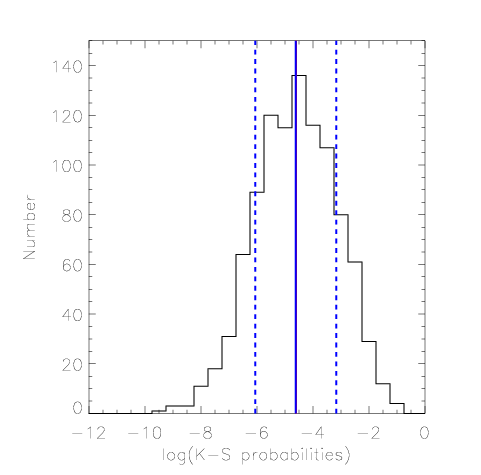}
\caption{The distribution of K-S probabilities that the combined protostellar lumionsity function of the high mass SF clouds are from the same parent distribution as the low mass SF clouds.  Shown are {\it log}(probability) from each of the 1000 Monte Carlo realizations.  The vertical blue bar shows the mean and vertical blue dashed lines show $\pm$ 1$\sigma$.}
\label{fig:fig10}
\end{figure}

\begin{figure}
\centering
\includegraphics[width=0.8\textwidth]{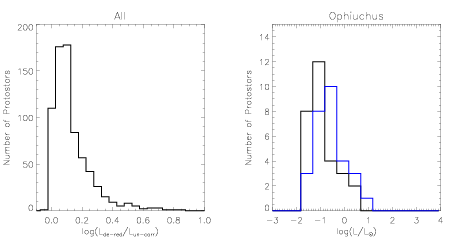}
\caption{Left panel: Ratio of bolometric luminosity determined using Equations \ref{eqn:relationship} and \ref{eqn:flat} to the de-reddend bolometric luminosity for protostar candidates in all clouds in this survey.  Right panel: Uncorrected Ophiuchus protostar candidate luminosity function (black) and de-reddened protostar candidate luminosity function (blue) using the technique described in Section \ref{sec:reddening}.}
\label{fig:fig11}
\end{figure}

\begin{figure}
\centering
\includegraphics[width=0.8\textwidth]{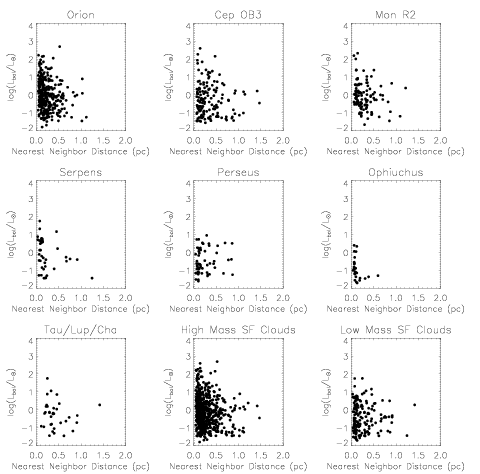}
\caption{Nearest neighbor distances (for 4$^{th}$ nearest neighbor YSO) and calculated bolometric luminosity for all protostar candidates.  The final panels show the protostars for Orion, Cep OB3, and Mon R2 and for Tau/Lup/Cha, Perseus, and Ophiuchus.  Orion, Cep OB3, and Mon R2 each have protostar candidates above 100 L$_{\odot}$ and show sources at larger nn4 distances only at lower luminosities.  The low mass SF clouds do not show this relationship individually or in the combined plot.}
\label{fig:fig12}
\end{figure}

\begin{figure}
\centering
\includegraphics[width=0.8\textwidth]{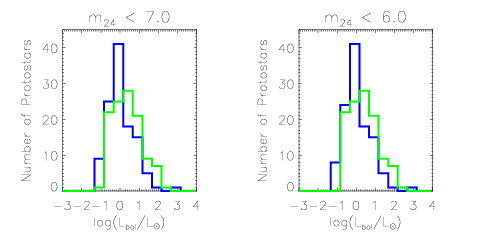}
\caption{Luminosity functions for the averaged high stellar density (green) and low stellar density (blue) contamination-subtracted protostars for $m_{24}$ $<$ 7 $mag$ (left) sources and for $m_{24}$ $<$ 6 $mag$ (right) sources.  The median K-S probability that the high and low stellar density luminosity functions are from the same parent distribution is 0.0013 for the $m_{24}$ $<$ 7 sources and 0.0015 for the luminosity functions of sources with $m_{24}$ $<$ 6.}
\label{fig:fig13}
\end{figure} 

\begin{figure}
\centering
\includegraphics[width=0.8\textwidth]{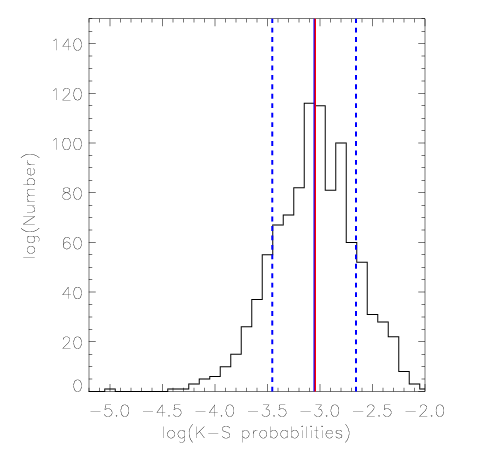}
\caption{Distribution of K-S probabilities that the populations of protostars in high and low stellar density in Orion are from the same parent distribution for each of the 1000 Monte Carlo realizations.  The median is shown as a vertical red bar, mean as a vertical blue bar, and $\pm$ 1$\sigma$ from the mean is shown as vertical blue dashes.}
\label{fig:fig14}
\end{figure}

\begin{figure}
\centering
\includegraphics[width=0.8\textwidth]{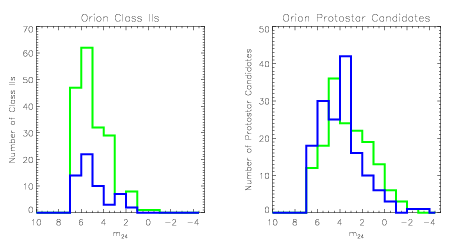}
\caption{Histograms of 24 $\mu$m detections for Orion high stellar density (green) and low stellar density (blue) YSOs.  Orion disk sources within $D_{c}$ of a protostar candidate in high or low stellar density (left panel) show similar distributions of 24 $\mu$m detections for sources in high and low stellar density regions and have a K-S probability of 0.61 for $m_{24}$ $<$ 7 sources.  Distributions of $m_{24}$ for the Orion protostar candidates in high and low stellar density (right panel) do not show similarities, and have a K-S probability of  0.0089 (at $m_{24}$ $<$ 7) that they are from the same parent distribution.}
\label{fig:fig15} 
\end{figure}

\begin{figure}
\centering
\includegraphics[width=0.8\textwidth]{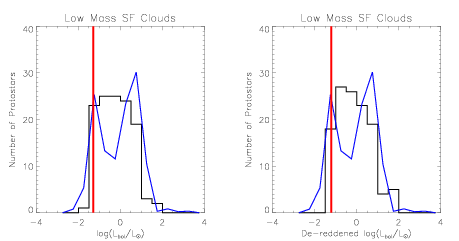}
\caption{Luminosity functions for the low-mass SF clouds from this work (black) and the luminosity function from the episodic accretion model (model 5) from \cite{2010ApJ...710..470D} (blue).  Shown are the uncorrected luminosity functions and the de-reddened luminoisty functions with the uncorrected and de-reddened $L_{cut}$, respectively (see Section \ref{sec:reddening}).}
\label{fig:fig16} 
\end{figure}

\begin{figure}
\centering
\includegraphics[width=0.8\textwidth]{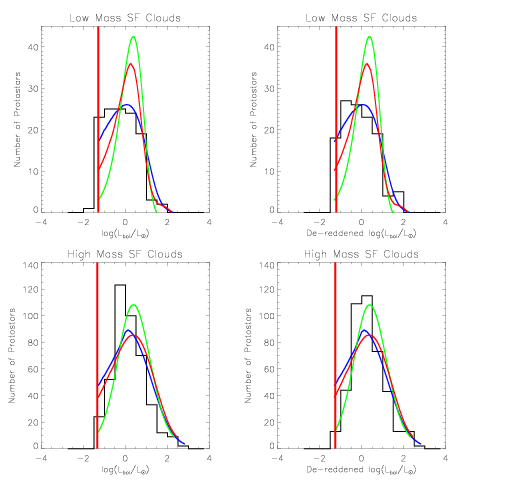}
\caption{Luminosity functions for the combined high mass SF clouds (Orion/Cep OB3/Mon R2) and the combined low mass SF clouds (Serpens/Perseus/Ophiuchus/Taurus/Lupus/Chamaeleon) are shown as histograms in black.  These are compared with models from \cite{2011ApJ...736...53O} including the competitive accretion model (blue), two-component turbulent core model (green), and turbulent core model (red).  The low mass SF clouds are compared with models which use an upper mass limit of 3 M$_{\odot}$, and the high mass SF clouds are compared with models corresponding to an upper mass limit of 10 M$_{\odot}$.  The left two panels show the uncorrected luminosity functions with uncorrected $L_{cut}$ (vertical red line).  The right two panels show the de-reddened combined luminosity functions and the de-reddened $L_{cut}$ (vertical red line).}
\label{fig:fig17} 
\end{figure}

\begin{figure}
\centering
\includegraphics[width=0.8\textwidth]{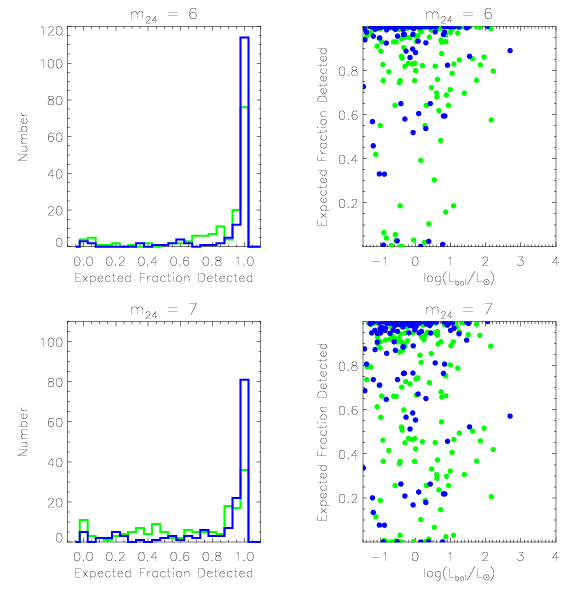}
\caption{The expected fraction of 24~$\mu$m sources recovered as a function of the confusion due to nebulosity and crowding for the Orion clouds.  We use the RMEDSQ technique of Megeath et al. (in prep.) to determine the expected fraction of sources detected with magnitudes equal to the cutoff magnitude.  We show the distribution of the fraction of detected sources in high stellar densities (green) and low stellar densities (blue) regions expected.  The top row of plots uses a cutoff magnitude $m_{24}$ = 6 $mag$, and the bottom row of plots use $m_{24}$ = 7 $mag$.  The left column of plots show the distribution of the expected fractions for all of the protostar candidates.  The right column of plots show the expected detection fraction at the cutoff vs. the luminosity of the protostar candidate.  We find that the $m_{24}$ = 6 $mag$ cutoff has a median expected detection fraction of 0.995 or 0.973 for the high stellar density protostar candidates and 0.999 for the low stellar density protostar candidates, and the $m_{24}$ = 7 $mag$ cutoff has a median expected detection fraction of 0.935 or 0.804 for the protostar candidates in high stellar density regions and 0.981 for the protostar candidates in regions of low stellar density.}
\label{fig:fig18}
\end{figure}

\end{document}